\documentclass[11pt,a4paper]{article}
\usepackage{fullpage}
\usepackage[utf8]{inputenc}
\usepackage{amsmath}
\usepackage{amsfonts}
\usepackage{amssymb}
\usepackage{authblk}
\usepackage{hyperref}
\usepackage[toc]{appendix}
\usepackage{graphicx}
\usepackage{rotating}
\usepackage{longtable}
\usepackage{booktabs}

\newcommand{\appendixnumberline}[1]{Appendix\space}

\providecommand{\keywords}[1]
{
	\small	\quad
	\textbf{Keywords --} #1
}

\let\oldappendix\appendix
\makeatletter
\renewcommand{\appendix}{%
	\addtocontents{toc}{\let\protect\numberline\protect\appendixnumberline}%
	\renewcommand{\@seccntformat}[1]{Appendix~\csname the##1\endcsname\quad}%
	\oldappendix
}
\makeatother

\begin{document}
\title{Department-level comparison of universities' scientific output: A bibliometric study of Greek universities}
\author{Ioannis Z. Koukoutsidis\thanks{This work was partly conducted while the author was employed at the Hellenic Authority for Higher Education (HAHE). The work is entirely based on publicly available data and no internal HAHE or university data were used.}}
%\affil{Hellenic Telecommunications and Post Committee, 60 Kifissias Avenue, 15125 Maroussi, Athens, Greece}
%\author[2]{Ioannis Vikas}
%\affil[2]{Hellenic Authority for Higher Education, 1 Aristidou \& 2 Evripidou Str., 10559 Athens, Greece}

\maketitle
\begin{abstract}
Although bibliometric studies for the assessment of scientific output at university-level are relatively common, data for performance at department-level rarely exist. In this paper we develop a methodology and tools for conducting department-level assessments, and conduct for the first time a complete bibliometric study of scientific output of all university departments in Greece. The study is based on data from Scopus about the number of scientific publications and respective citations of all faculty members in each department for the period 2017-2021. Code scripts were developed using R to query the Scopus database and automatically generate statistics, as well as an online application to view the results. The results reveal interesting facts about the scientific impact of greek university departments, which vary between universities and thematic areas. The developed tools can be used for continuous monitoring and evaluation of university departments worldwide. 
\end{abstract}

\keywords{university departments, bibliometric analysis, Scopus, R code}

\section{Introduction}
The issue of quality in higher education is gaining interest among education authorities and universities worldwide. 
One aspect of interest is the quality of research conducted in universities, which is predominantly reflected in the number of scientific publications and their respective citations. Several research studies exist that evaluate performance across universities and disciplines (e.g. see \cite{moed2006bibliometric,cancino2017bibliometric}), and many university rankings include citation analysis as one of the assessment criteria\footnote{Well-known rankings are the Times Higher Education World University Rankings, the QS World University Rankings, the CWTS Leiden Ranking, and U-Multirank.}. Relevant statistics are monitored by universities as well as education authorities to assess the scientific impact of university research at institutional, national and international level. These are subsequently used to find strong and weak points of university research, improve resources and procedures, allocate funding, and  take policy decisions at institutional or authority level. In Greece, apart from monitoring and policy making purposes, the Hellenic Authority for Higher Education (HAHE) and the Ministry of Education and Religious Affairs use the number of publications and respective citations as one of the criteria to determine the distribution of the 20\% of annual state subsidies between universities, in a recently published regulation \cite{regulation22}.

Among the various citation databases that exist, the most well known for bibliometric analysis are Scopus, Web of Science (WoS) and Google Scholar.\footnote{Microsoft Academic was another well-known search engine, which was discontinued in 2022. Other well known scholar databases are Semantic Scholar, AMiner, CrossRef and Dimensions AI.} All automatically collect data from publishers (books, journals, conference proceedings, abstracts) and build publication and citation indices. Scopus and WoS have paid access analytics environments (scival.com for Scopus, InCites.clarivate.com for WoS), where the data are analyzed and useful impact statistics are derived. Google Scholar is a very useful open source of data, particularly for search and reference purposes, but it significantly lags behind Scopus and WoS in terms of data analysis and processing for producing impact statistics. 

The Scopus database was used for this study. Compared to WoS, Scopus has a higher coverage of journals published outside the US, and also covers more journals in different disciplines.\footnote{At the time of writing, Scopus contained 81 million documents.}. It also applies quality criteria for indexing sources; namely, peer-reviewing of publications and that a new title maintains some quality standards or is part of a series (e.g. an annual conference as opposed to an one-off event). The establishment of quality criteria trades-off quality for coverage; compared to Google Scholar, Scopus has a smaller coverage, which is also reflected in the citation metrics (i.e. results in much smaller numbers of publications and respective citations compared to Google Scholar, which also includes a lot of non-reviewed content). Nevertheless, in terms of bibliometric analysis, Scholar offers the most possibilities. 
It is also the database of choice for the calculation of quality indicators in the Greek regulation for the assessment of university performance \cite{regulation22}. Interested readers are referred to \cite{harzing2016google} for a more in-depth comparison of these databases.

Scopus has great capabilities for the generation of impact statistics at {\em university level}, including comparisons between universities. However, when one is interested at a deeper analysis at {\em department level} inside universities, citation statistics rarely exist. This can be attributed to two reasons: first, many authors do not add department affiliations in their papers; second, department affiliations are in many cases not indexed by the databases, or indexed as part of the university. For example, out of nine Electrical and Computer Engineering Departments in Greece, Scopus indexes only one (!), whereas out of four Chemical Engineering Departments, Scopus search returns a reference to the National Technical University of Athens (one of the universities hosting such department), and two references to specialized chemical engineering institutes in Greece.

Nevertheless, the monitoring of citation statistics at department level can be very useful, not only for universities and education authorities, but also for prospective undergraduate and post-graduate students and their families. Apart from a few highly-reputable university departments, the performance of other departments is largely unknown to the broader public. Particularly for prospective students who finish secondary education and first enter an academic environment, having access to such information could play a significant role in their choice of university department.

This paper proposes to fill this gap by providing a methodology and tools for department-level citation analysis. As a case study, we perform, for the first time, a citation analysis of {\em all} university departments in Greece, and present performance comparisons of departments in each university, as well inter-university comparisons between departments with the same thematic area. Data about the number of publications and respective calculations are collected using the Scopus database. Results are presented in the paper; given their large volume, a web application has also been created where interested readers can easily browse through the results for all universities and departments. Besides the value of the results per se, the major goal of the paper is to reveal the importance of such statistics, and to call upon universities and education authorities to establish procedures for generating and updating such results. As soon as a systematic data collection procedure is established, research impact statistics can be generated with relatively little effort, offering valuable knowledge.

In previous work, the authors in \cite{altanopoulou2012evaluation} performed an evaluation of a subset of Greek university departments and their faculty members. 93 university departments and 3358 faculty members were included in this study, and citation statistics were collected based on Google Scholar. In \cite{Pitsolanti2018}, the authors investigated 50 Greek Science and Engineering University Departments and 1978 academics, producing statistics about the number of papers, citations, h-index and i10-index using Google Scholar. Most other work from other countries is also limited in scope, although advanced statistical analysis is sometimes performed. In \cite{krampen2008evaluation}, the author performed a bibliometric publication and citation analysis for a University Department of Psychology, also comparing the results with self-and colleagues-perceptions for some faculty members. In \cite{kotsemir2017measuring}, a bibliometric study of three departments in a Russian university was conducted using Scopus, also making comparisons between ``new" and ``old" staff members. Our study comprises all departments and faculty members in Greece (425 university departments and 9838 faculty members), and is by far the most complete and systematic study done so far in this field.
\section{Methodology and metrics}
For this research, all data for publications and respective citations have been retrieved from the Scopus database.
In comparison with Web of Science and Google Scholar, Scopus has a publicly searchable system of unique author and affiliation identification, which makes it particularly suitable for bibliometric analysis. It also has an Application Programming Interface (API) suitable for programmatic access to data, which was also exploited in the paper to automatically retrieve publications for groups of authors.

The most arduous part of the work was to collect the unique identifier of each author (so called ``Scopus ID''), which is then used to retrieve bibliometric information. This task was done manually from December 2021 until August 2022 (the actual work took about 6 person months). It was a very laborious process, not only because it demanded a manual search for Scopus profiles of 9838 greek faculty members, but also because it was more than a simple search: besides searching for the Scopus ID, it also involved three major tasks: a) sorting out duplicate names (i.e. Scopus profiles with the same name, frequently occurring for faculty members with common surnames), b) merging profiles belonging to the same author (Scopus automatically creates a different profile when it cannot match authorship to an already created profile --- this {\em may} happen when an author changes affiliation, name, co-authors, or publishes in a different subject area). This also occurred frequently (in two cases, the authors had 6 different profiles in Scopus!). Throughout the work, 1131 merge requests were sent to Scopus, after verifying (through the author list of publications from the university or personal website) that they indeed belonged to the same author, and c) correcting erroneous previous merges (so-called unmerging): although this was relatively rare, in some cases we found that a faculty member's profile was ``contaminated'' with publications from another author. This was even observed for authors from totally different fields (e.g. economics and medicine). For these cases, Scopus' option for requesting removal of publications from the author's profile was used, after confirming that the publications did not belong to the searched faculty member. 

A Scopus profile was not found for 1158 faculty members (11.77\%), despite best efforts (search for the author with different name options, or using the author's publications to search). As anticipated, this was encountered mostly for professors in social sciences, humanities, law, and arts. Most of the authors in these disciplines publish in Greek language, the vast majority of which is not covered by Scopus. Although these departments were also included in the study, results should be seen with great caution and not used for conclusions about the departments' performance.

After collecting the Scopus IDs, a collective search for the authors' bibliographic data and calculated statistical metrics was performed, by exploiting the Scopus API and creating scripts in the R language using the bibliometrix package \cite{bibliometrix} to retrieve the data and generate statistics. The created scripts have been published on GitHub in the form of R functions and are freely available to everyone \cite{Koukoutsidis_Bibliostats_GitHub_repository_2022}.

The metrics used to evaluate performance of university departments are the number of papers by the department's faculty members and the number of papers per faculty member in the last 5 years, the number of citations to these papers and the resulting citations per faculty member. The 5-year period is dictated in \cite{regulation22} and puts more emphasis on recent achievements and corresponding impact. The results in this study refer to the period 2017--2021.

For each university department we automatically retrieve the list of publications of all faculty members in the above period and the number of citations to these papers, and calculate the above statistics. Duplicate publications (i.e. publications co-authored by more than one faculty members in the same department) are excluded, whereas we permit self-citations.
\section{Results}
Appendix~\ref{appendixA} shows the list of greek higher education institutions covered in this study -- in decreasing faculty size -- and their common abbreviations. These constitute all greek higher education institutions, except for military academic institutions and ecclesiastical schools.\footnote{Although they also belong to Greek Higher Education, military and ecclesiastical schools constitute separate categories and were excluded from the study. Additionally, military schools do not publish the names of faculty members online.} 

In all results we maintain the official name of greek faculty, which is TRS (Teaching and Research Staff). This includes tenured faculty at all levels (Professor, Associate Professor, Assistant Professor, and Lecturer\footnote{The Lecturer position is deprecated in Greek Higher Education, as all new employments start from Assistant Professor, but still some persons exist having this title.}), as well as probationary assistant professors. We exclude temporary faculty members, such as adjunct professors, as well as other collaborators (e.g. students, post-docs).\footnote{We only consider TRS members, both because of the difficulty of getting data for temporary staff and students, as well as because it would distort the publication ratios. In terms of number of publications the error is expected to be small, since papers produced by temporary staff, students and post-docs usually include permanent staff as co-authors.}. 

Note also that the same scientific divisions in one university may be called a ``Department'', and in another a ``School'', or a ``Faculty'' (e.g. the Physics division is called a School in AUTH, whereas in NKUA it is called a Department). This is a matter of naming convention of each University, and in practice it is the same.

There are two categories of results: a) comparisons between departments in the same university, and b) comparisons between departments in the same thematic area (subject) from different universities. We rank departments using citations per TRS member and citations per paper. The first metric shows roughly the impact (in terms of citations) of the average TRS member of the department, whereas the second the average impact of papers produced by the department. In many cases, we find differences in the ranking of departments when one or the other metric is considered; this occurs often in disciplines where there are usually a lot of co-authors per paper (e.g. medicine, biology). Therefore for completeness it is deemed as important to consider both metrics.

\begin{sidewaysfigure}[ht]
	\includegraphics[width=\textwidth]{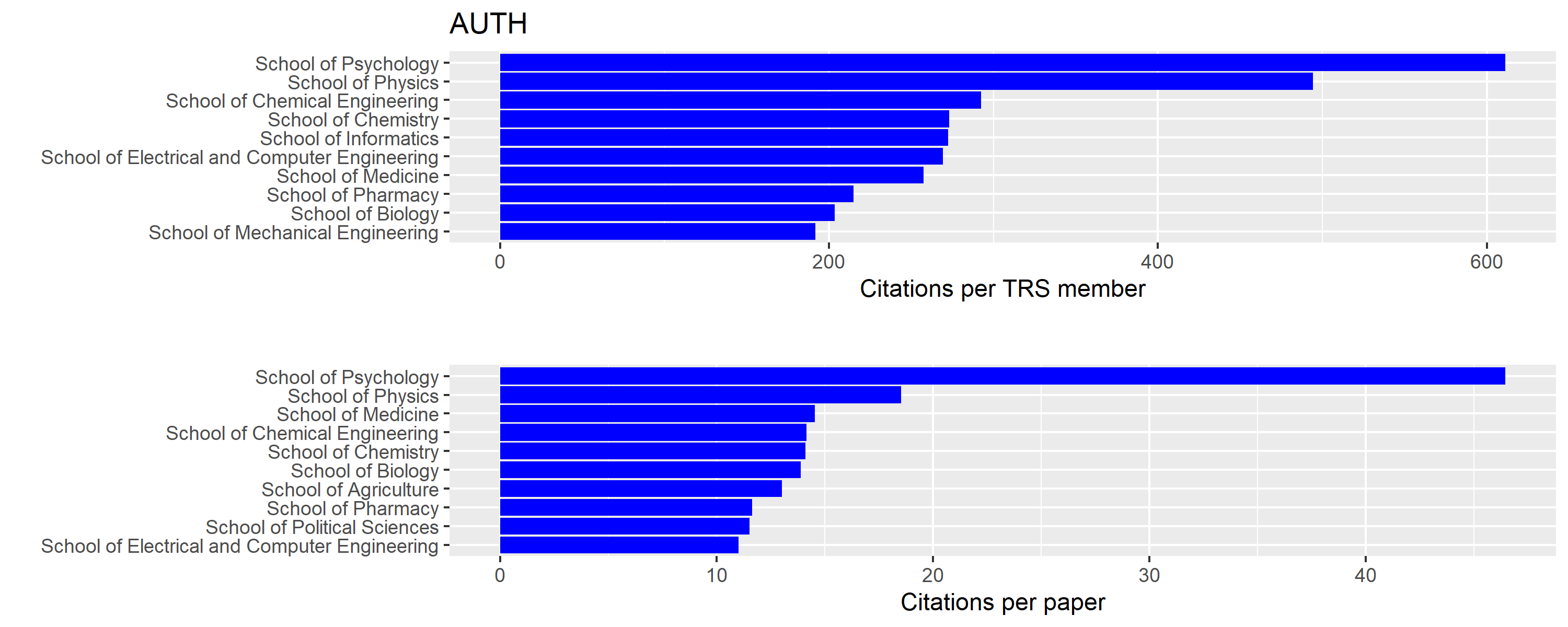}
	\caption{Ranking of top-10 departments (AUTH), based on publications in 2017--2021}
	\label{ranking:AUTH}
\end{sidewaysfigure}

\begin{sidewaysfigure}[ht]
	\includegraphics[width=\textwidth]{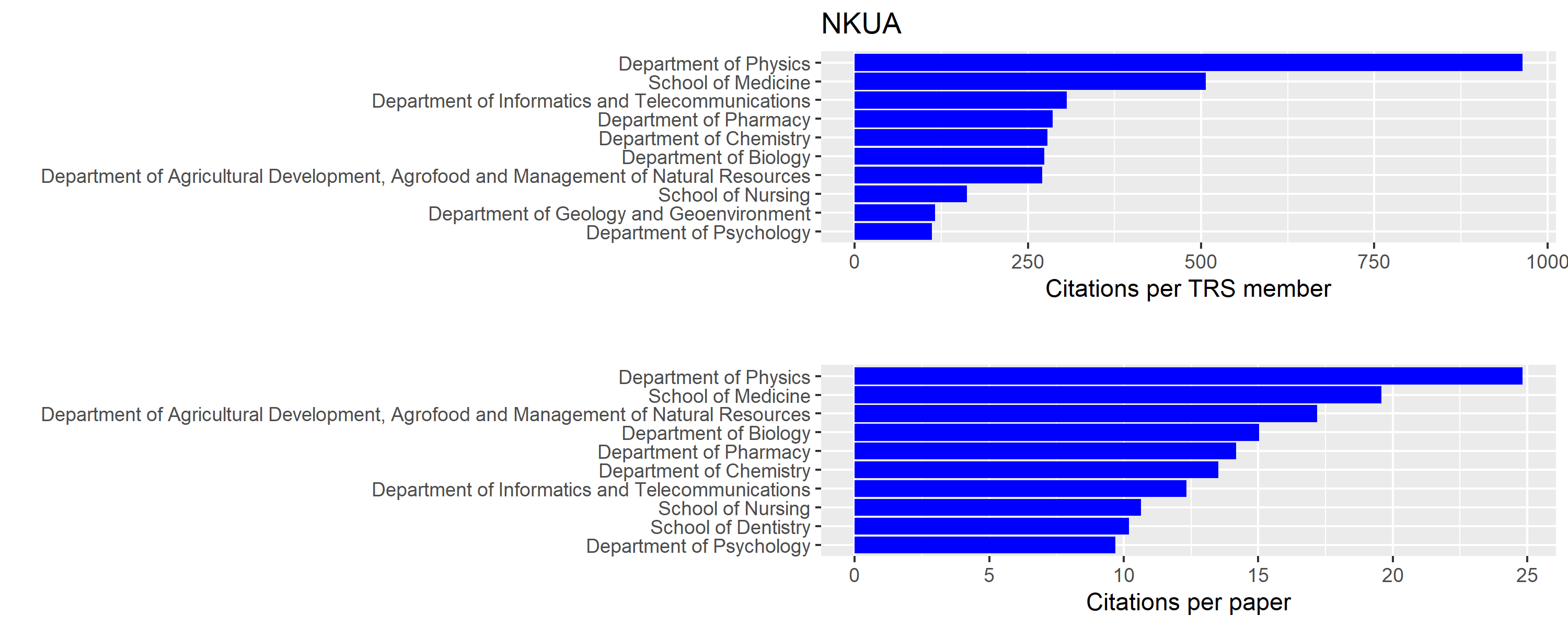}
	\caption{Ranking of top-10 departments (NKUA), based on publications in 2017--2021}
	\label{ranking:NKUA}
\end{sidewaysfigure}

Due to lack of space, in this section we only present a subset of the results derived in the study and refer to Appendix~\ref{appendixB} for the full table of results. Figures \ref{ranking:AUTH}--\ref{ranking:NKUA} show the ranking of departments within the same university, for the two biggest universities in Greece: AUTH and NKUA. It is reminded that all results refer to publications in years 2017--2021. In AUTH, it was surprising to find the School of Psychology ranking first both in citations per TRS member and citations per paper, above the School of Physics\footnote{Physics departments are generally high in rankings; during the study, we found that many members of these departments are also involved in large-scale collaborations, such as the ATLAS Collaboration in CERN, which results in extremely high publication rates. Large-scale collaborative publications were also found in medicine, e.g. CovidSurg.}. In NKUA, the top-2 is made of the Department of Physics and the School of Medicine, whereas the Department of Psychology is 10th. The departments that are in the top-10 according to citations per TRS member are usually also in the top-10 according to citations per paper, but this is not always the case; for example, in AUTH the School of Agriculture is not even in the top-10 of citations per TRS member, however it is 7th according to citations per paper. In NKUA, the Department of Dentistry gets in the top-10 of citations per paper, whereas it is absent from the top-10 of citations per TRS member.

\begin{sidewaysfigure}[ht]
	\includegraphics[width=\textwidth]{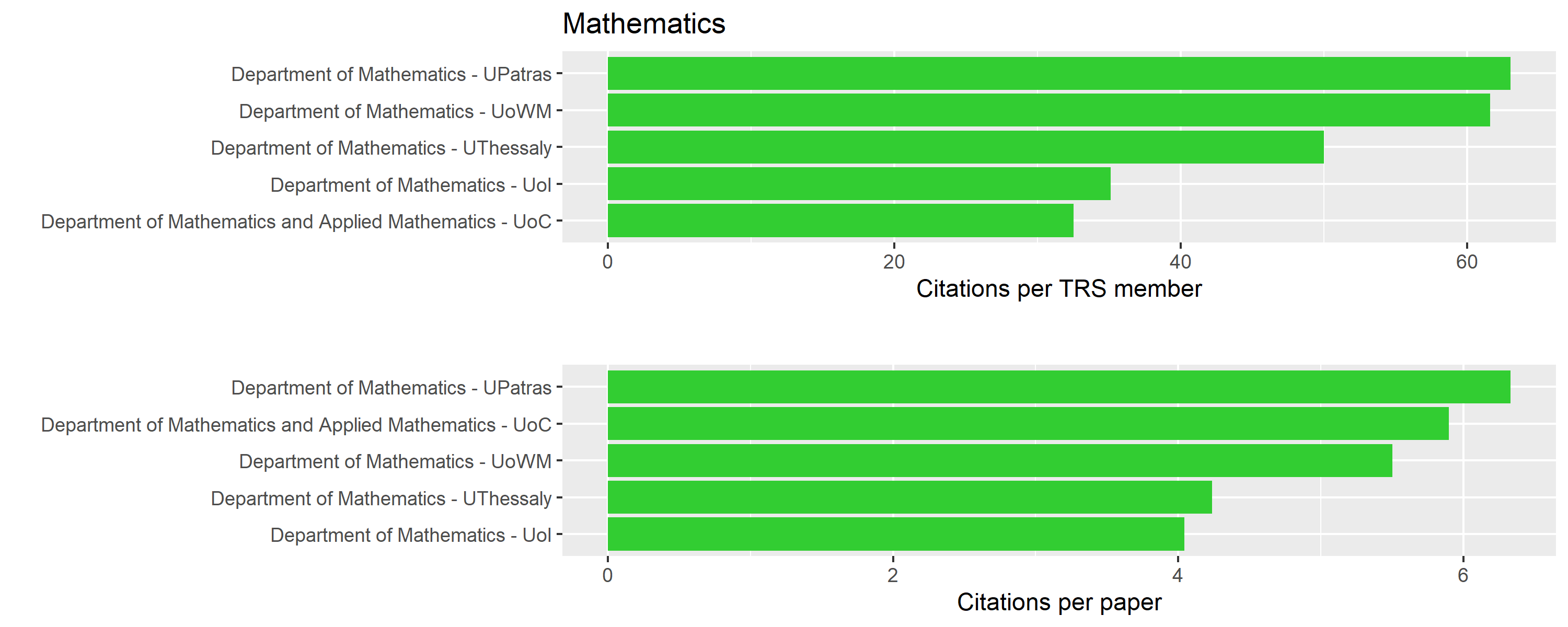}
	\caption{Ranking of top-5 departments in mathematics, based on publications in 2017--2021}
	\label{ranking:math}
\end{sidewaysfigure}

\begin{sidewaysfigure}[ht]
	\includegraphics[width=\textwidth]{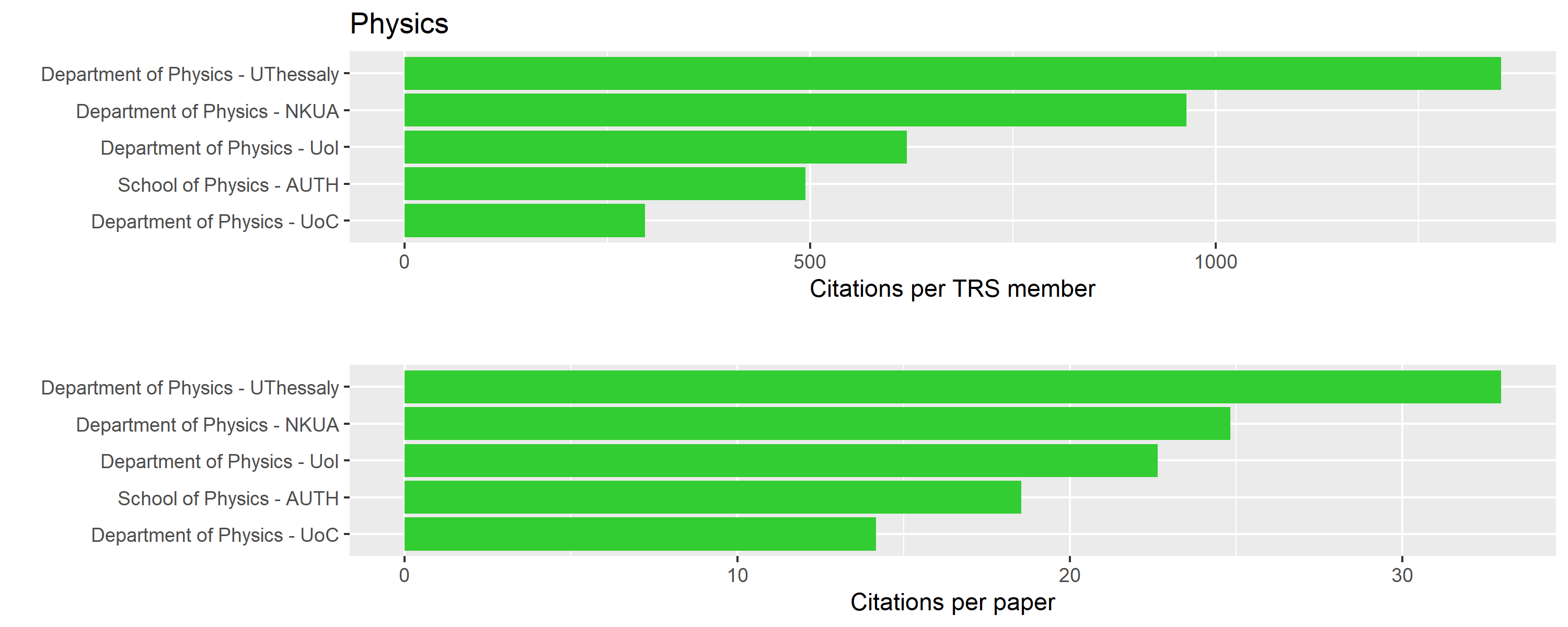}
	\caption{Ranking of top-5 departments in physics, based on publications in 2017--2021}
	\label{ranking:phys}
\end{sidewaysfigure}

\begin{sidewaysfigure}[ht]
	\includegraphics[width=\textwidth]{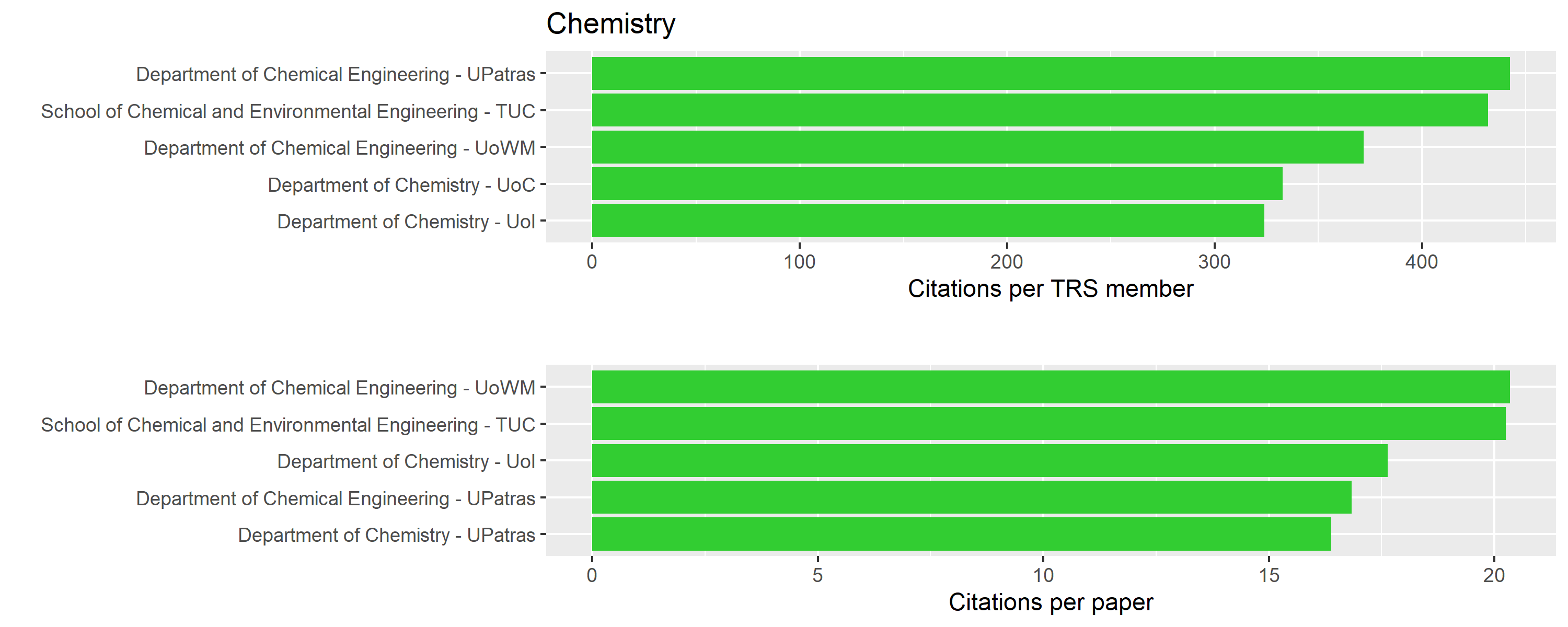}
	\caption{Ranking of top-5 departments in chemistry, based on publications in 2017--2021}
	\label{ranking:chem}
\end{sidewaysfigure}

\begin{sidewaysfigure}[ht]
	\includegraphics[width=\textwidth]{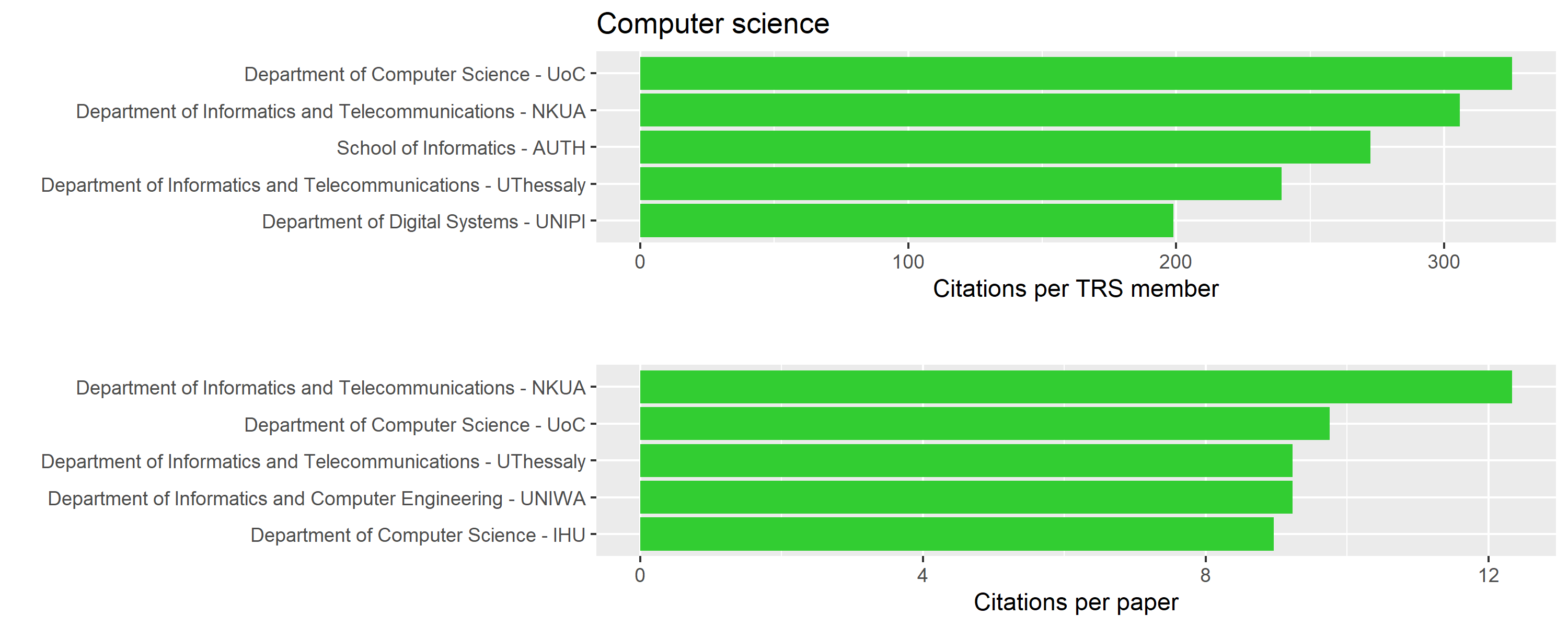}
	\caption{Ranking of top-5 departments in computer science, based on publications in 2017--2021}
	\label{ranking:comp}
\end{sidewaysfigure}

\begin{sidewaysfigure}[ht]
	\includegraphics[width=\textwidth]{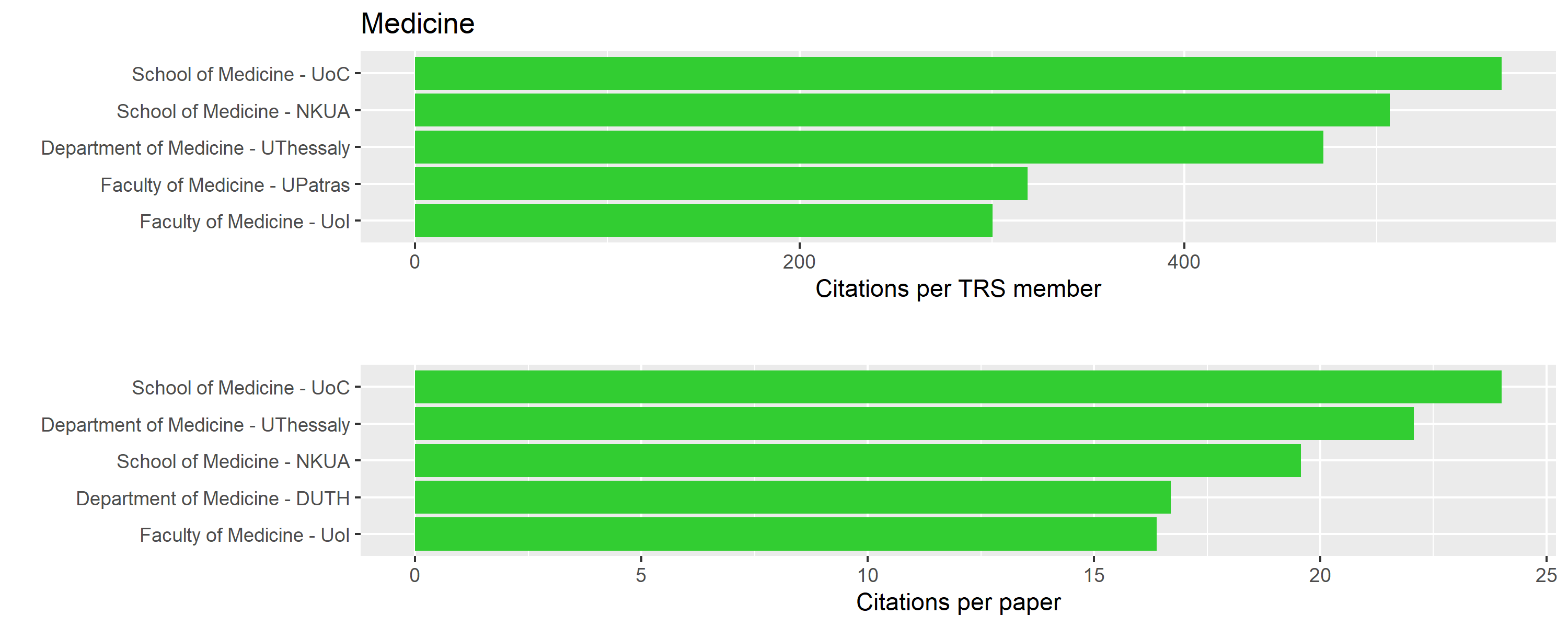}
	\caption{Ranking of top-5 departments in medicine, based on publications in 2017--2021}
	\label{ranking:med}
\end{sidewaysfigure}

\begin{sidewaysfigure}[ht]
	\includegraphics[width=\textwidth]{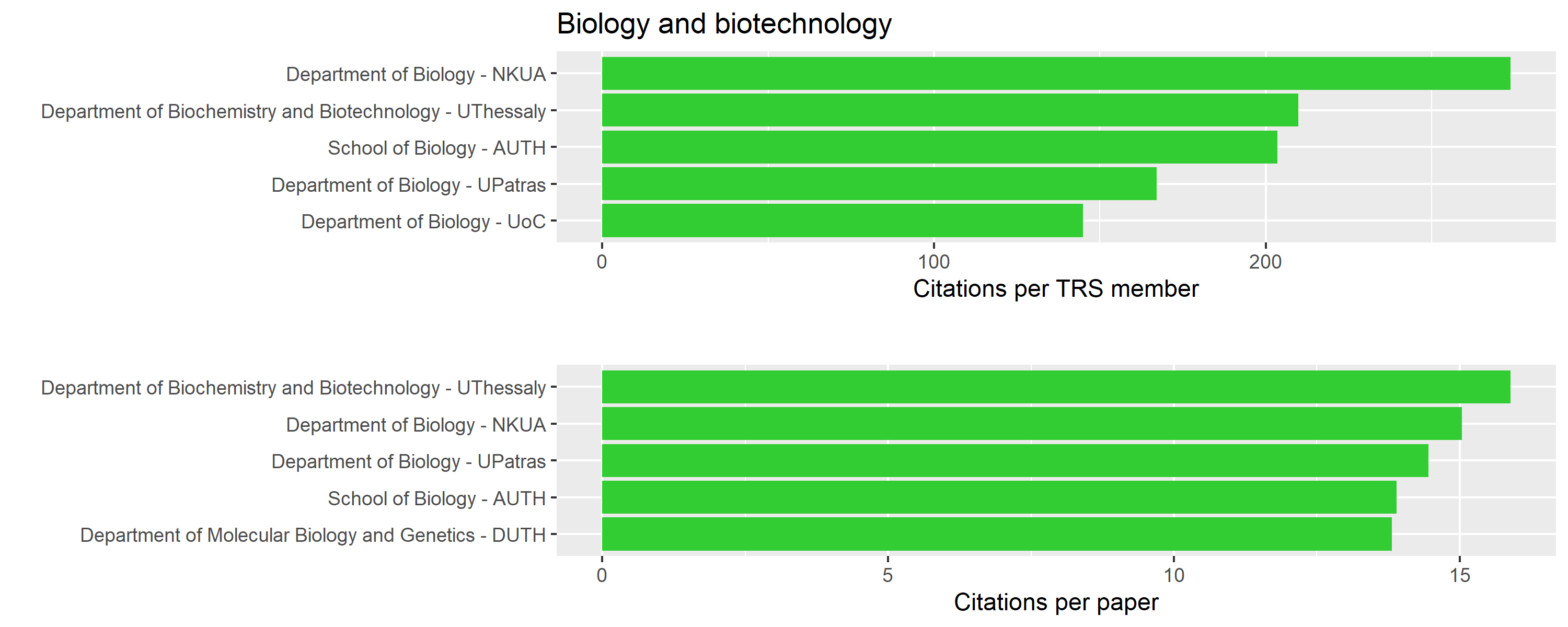}
	\caption{Ranking of top-5 departments in biology and biotechnology, based on publications in 2017--2021}
	\label{ranking:bio}
\end{sidewaysfigure}

\begin{sidewaysfigure}[ht]
	\includegraphics[width=\textwidth]{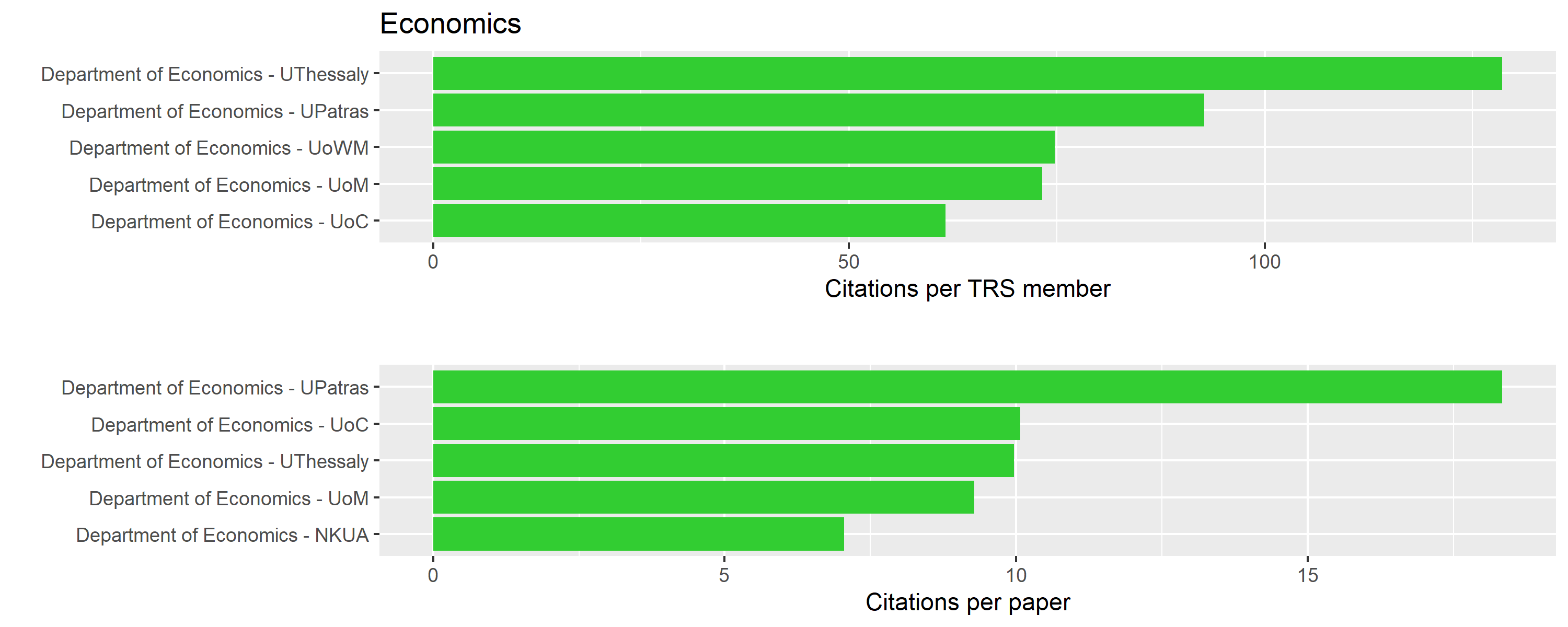}
	\caption{Ranking of top-5 departments in economics, based on publications in 2017--2021}
	\label{ranking:econ}
\end{sidewaysfigure}

\begin{sidewaysfigure}[ht]
	\includegraphics[width=\textwidth]{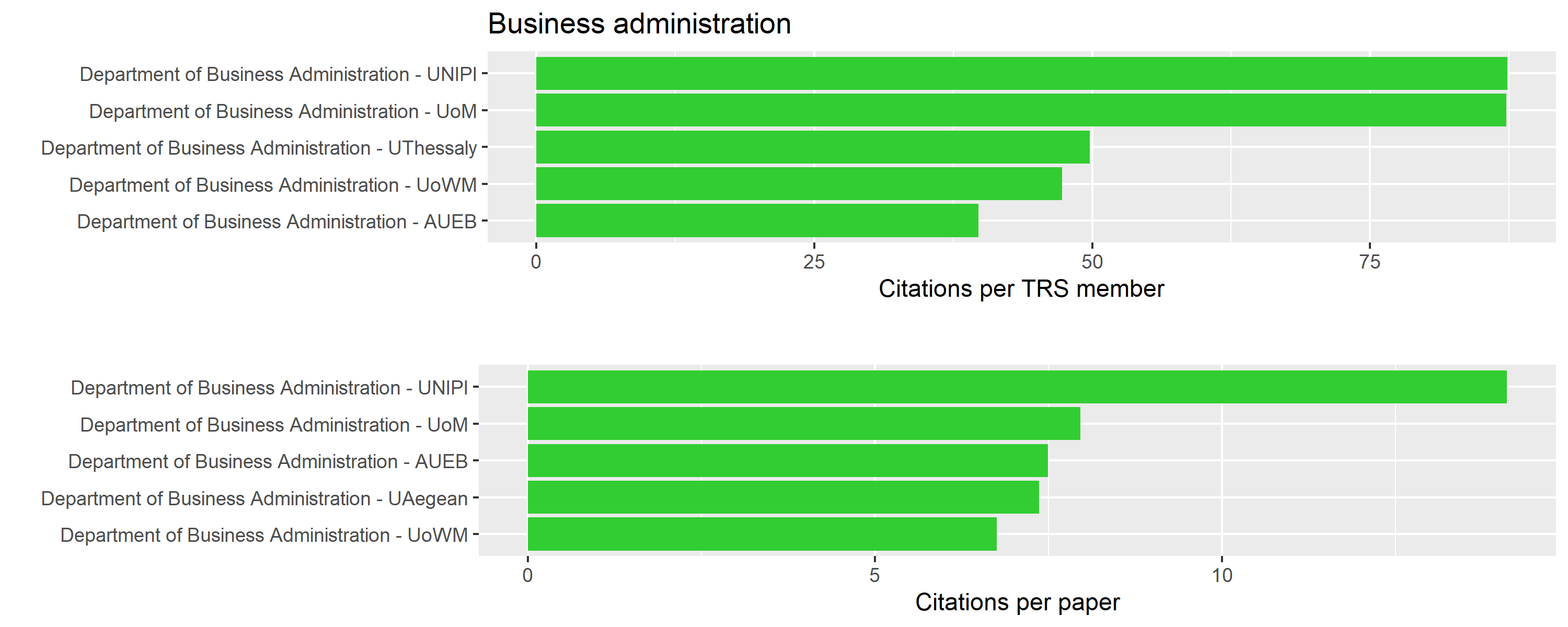}
	\caption{Ranking of top-5 departments in business administration, based on publications in 2017--2021}
	\label{ranking:bizz}
\end{sidewaysfigure}

In Figures~\ref{ranking:math}--\ref{ranking:bizz} we show the top-5 university departments in Greece in 8 different scientific disciplines: math\footnote{We have not included the School of Applied Mathematical and Physical Sciences of NTUA in this comparison, since it is an interdisciplinary department.}, physics, chemistry, computer science, medicine, biology, economics and business. This is the type of result that is perhaps most interesting, as it shows how a department stands compared to other departments in the same area. 
We first see that both metrics vary greatly across disciplines: the best math department has a little over 60 citations per TRS member and a little over 6 citations per paper, whereas in physics the respective metrics for the best department are over 1250 and over 30, respectively. This is the most extreme difference, but in general one can see many differences across disciplines. On this respect, it is also interesting to notice that the citations per paper metric is much more stable across scientific disciplines, whereas citations per TRS member varies more significantly (from a few tens in math and business departments, to a few hundreds in other disciplines).

In general, we also see the same pattern as in intra-university comparisons: top departments according to citations per TRS member are usually also top according to citations per paper, with some notable exceptions: for example, in computer science, we see that the School of Informatics of AUTH is third according to citations per TRS member, whereas it is not in the top-5 according to citations per paper. Instead we see other less well known universities, like UNIWA and IHU entering the top-5. 

The ranking according to number of citations per TRS member also tends to favor small-size departments. For example, the Departments of Mathematics in UoWM and UThessaly have only 5 TRS members, which is significantly lower than UPatras (23), UoC (33) and UoI (19). In consequence, they are ranked higher according to this metric, than when considering the impact of the papers alone. This is generally observed across disciplines. Indeed, in small-size departments it is less likely to have many joint publications between TRS members. Nevertheless, it is nice to see many small-size departments ranking in the top-5 in both metrics, which is certainly also attributed to the quality and efforts of their TRS.

Detailed results can be found in Appendix~\ref{appendixB} for all departments covered in this study. In each university, departments are listed in decreasing order of citations per TRS member. Apart from the two major metrics, we also present in this table the number of TRS members, the number of papers in 2017--2021, number of citations to these papers, and number of papers per TRS member.

\section{Web application}
To further increase the visibility of our results and make them more accessible to the wider public, a web application was created using R Shiny \cite{shiny}, which is available online in both greek and english \cite{Koukoutsidis_biblio_shinyapp}.

\begin{figure}[!htb]
	\centering
	\includegraphics[width=\textwidth]{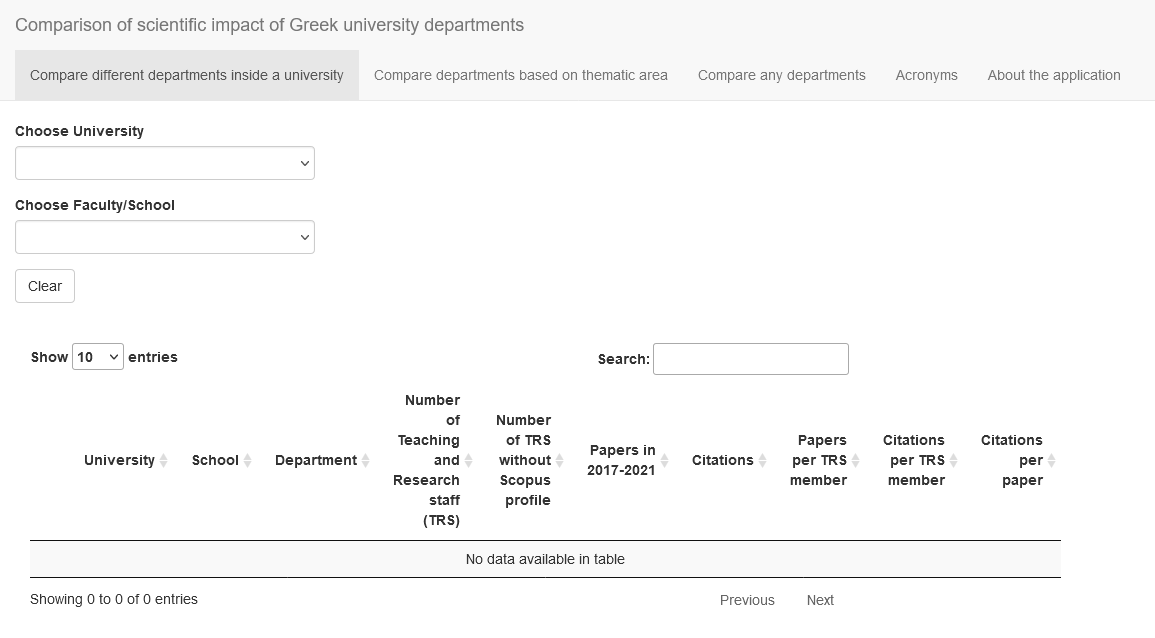}
	\caption{Screenshot of initial page of web application, December 2022}
	\label{screenshot}
\end{figure}

Figure~\ref{screenshot} shows a screenshot of the initial page of the application. Three main comparison functionalities were implemented, each at a different tab of the app: a) comparison between departments in a university, or between departments in a faculty/school of the university (where such subdivision exists), b) comparison based on thematic area, where the user can type search terms in a box and a list of related departments from all greek universities is shown (there is also the possibility to exclude some departments from the list), and c) comparison between any departments, where the user can select and compare up to five university departments in any area.

All results are shown in table format. The same results are shown as in Appendix~\ref{appendixB} of this paper, with the addition of the number of TRS members in each department without a Scopus profile. The user also has the option to rank the results in increasing or decreasing order, according to any variable. The application is responsive and can be used in both mobile and desktop devices.
\section{Conclusions and discussion}
This paper has presented a large-scale bibliometric study of university departments in Greece, based on the analysis of publications and respective citations of permanent faculty members in the departments for the period 2017--2021. The study covered 425 departments (practically all higher education departments) and 9838 faculty members, and is by far the most complete study done so far in this field.

It is worth noting that no proprietary data or internal university data were used for this study. All data used were publicly available, and can be retrieved by anyone. The Scopus database was used as data source, as it is most suitable for bibliometric analysis, and offers a convenient API for programmatically querying the database. Although Scopus currently seems to be the golden standard for many bibliometric studies, it still significantly lacks in coverage of publications in languages other than english, as well as in social, humanitarian, and legal sciences. In many cases, we found authors in these fields having a large number of publications and citations in Google Scholar, but not indexed in Scopus. In view of this under-reporting, it is not wise to use this tool for evaluation of authors or departments in these fields. Arts and architecture are also categories of their own, where scientific excellence is measured by other means, such as number of produced designs, artworks, number of exhibitions, etc.

The study revealed many interesting results, such as the surprisingly good performance of some psychology departments (as also discussed previously), or of some departments in nutrition and food science, and even physical education and sport science. Many good departments were found in small and less well-known universities, and even in departments with few permanent faculty members. Overall, it is useful to review all data carefully, since they vary significantly across universities and subject areas.

Regarding metrics, it was remarked that the number of citations per paper is generally a more stable metric than citations per TRS member, as the latter varies more significantly across departments. Thus, although the citations per TRS member metric is more frequently used in evaluations (it is also used in \cite{regulation22}), it is worth using both metrics.

Apart from the results, the study made contributions to software for bibliometric analysis, creating code to automatically collect data and generate statistics that have been published as open source in \cite{Koukoutsidis_Bibliostats_GitHub_repository_2022} and can be freely used by other researchers. Besides code for Scopus, a program was also developed for scraping a Google Scholar institution (e.g. university) page and calculating statistics, which has also been published online.

Only basic statistics were derived in the paper, as it focused on covering all departments rather than performing a more complex analysis. In future work, it would be interesting to study distributions of publications and citation numbers over TRS members in both small and large universities (i.e. what percentage of TRS members are responsible for the bulk of publications in each department), and how the results correlate with h-index values or other metrics. These results are not easy to conduct in large scale, and more advanced routines are needed.

A web application was also developed to facilitate browsing through the results and to make them available to the wider public. The application is static, i.e. it reads from the already derived results. A more evolved application could also be developed in the future, dynamically retrieving fresh data and updating the results. 

As the results for the performance evaluation of university departments in some period are by nature ephemeral, perhaps the biggest contribution of this work is to show that a large-scale bibliometric analysis at department level is possible. The work conducted here was particularly arduous, because it involved collecting Scopus Author IDs for the first time for all departments and involved a lot of ``tidying up'' of the Scopus database with necessary profile merges (and some unmerges). However, once this is done and the basic dataset is created, then maintaining and updating it is much easier. 

More importantly, this part of the work should be done by the departments and the faculty members themselves. Universities should advise their faculty members to maintain their Scopus profiles, doing necessary merges if they have more than one profile and review the contained publications in each profile to correct any errors. Additionally, the department web pages could be improved by systematically providing author IDs of their faculty members. During the study, it was disappointing to see that a relatively small number of departments had the Scopus IDs of their faculty members online on their webpages. To facilitate evaluation studies, education authorities could also mandate that all university departments provide the Scopus IDs (or other identifier that may be used in the future, e.g. ORCID) of their faculty members as part of the data they have to provide.  

Overall, the author believes that it is worth systematizing this process, and regularly providing results (e.g. annually).  It is by no means argued that publications and citations are the sole metrics that are important for evaluating performance, but they are likely to be correlated with good performance in general. If applied internationally, it could even facilitate comparisons of departments in the same thematic area from different countries, for which also there is hardly any evidence. Generally, such results can compensate -- to some degree -- for the general lack of public evidence of university performance at department level, and can bring great value to the academic community, education authorities and the wider public.   
\section*{Acknowledgement}
The author would like to acknowledge the help of Dr.~Ioannis Vikas from the Hellenic Authority of Higher Education in collecting Scopus Author IDs for some of the departments, as well as for useful discussions during part of this work.
\bibliography{ethaee_paper_arxiv}
\bibliographystyle{plainurl}

\clearpage
\appendix
\section{List of Higher Education Institutes}
\label{appendixA}
\begin{table}[h!]
	\centering
	\caption{List of Higher Education Institutions in Greece}
	\begin{tabular}{p{7cm}p{2cm}p{2.5cm}} 
		\toprule
		Instution name & Abbreviation & Number of TRS members \\  
\midrule
		Aristotle University of Thessaloniki & AUTH & 1628 \\ 
		National and Kapodistrian University of Athens & NKUA & 1584 \\
		University of Thessaly & UThessaly & 667 \\
		University of Patras & UPatras & 654 \\
		University of West Attica & UNIWA & 599 \\ 
		University of Ioannina & UoI & 526 \\ 
		Democritus University of Thrace & DUTH & 510 \\ 
		University of Crete & UoC & 448 \\
		National Technical University of Athens & NTUA & 404 \\
		International Hellenic University & IHU & 394 \\
		University of the Aegean & UAegean & 318 \\
		University of the Peloponnese & UoP & 296 \\
		University of Western Macedonia & UoWM & 210 \\
		Agricultural University of Athens & AUA & 199 \\
		Panteion University of Social and Political Sciences & Panteion & 199 \\
		University of Piraeus & UNIPI & 185 \\
		Ionian University & IONIO & 181 \\
		Athens University of Economics and Business & AUEB & 179 \\
		University of Macedonia & UoM & 171 \\
		Hellenic Mediterranean University & HMU & 168 \\
		Technical University of Crete & TUC & 117 \\
		Harokopio University of Athens & HUA & 70 \\
		Higher School of Pedagogical and Technological Education & ASPAITE & 47 \\
		Hellenic Open University & HOU & 44 \\
		Athens School of Fine Arts & ASFA & 40 \\		
		\bottomrule
	\end{tabular}
	\label{list_of_unis}
\end{table}
\clearpage

\section{Detailed results}
\label{appendixB}

% latex table generated in R 4.2.1 by xtable 1.8-4 package
% Sat Dec  3 17:29:56 2022
\begin{longtable}{p{4cm}p{1.5cm}p{1.5cm}p{1.5cm}p{2cm}p{1.5cm}p{1.5cm}}
	\toprule
	Department name & Number of TRS & Papers in 2017-2021 & Citations & Papers per TRS member & Citations per TRS member & Citations per paper \\ 
	\midrule
	\endhead
	\multicolumn{7}{l}{AUTH}\\
	\midrule
School of Psychology & 19 & 250 & 11615 & 13.16 & 611.32 & 46.46 \\ 
School of Physics & 63 & 1680 & 31132 & 26.67 & 494.16 & 18.53 \\ 
School of Chemical Engineering & 22 & 454 & 6433 & 20.64 & 292.41 & 14.17 \\ 
School of Chemistry & 61 & 1181 & 16656 & 19.36 & 273.05 & 14.10 \\ 
School of Informatics & 30 & 1123 & 8174 & 37.43 & 272.47 & 7.28 \\ 
School of Electrical and Computer Engineering & 47 & 1149 & 12654 & 24.45 & 269.23 & 11.01 \\ 
School of Medicine & 328 & 5806 & 84500 & 17.70 & 257.62 & 14.55 \\ 
School of Pharmacy & 20 & 369 & 4299 & 18.45 & 214.95 & 11.65 \\ 
School of Biology & 42 & 615 & 8546 & 14.64 & 203.48 & 13.90 \\ 
School of Mechanical Engineering & 28 & 537 & 5373 & 19.18 & 191.89 & 10.01 \\ 
School of Agriculture & 62 & 826 & 10764 & 13.32 & 173.61 & 13.03 \\ 
School of Geology & 38 & 502 & 4594 & 13.21 & 120.89 & 9.15 \\ 
School of Veterinary Medicine & 72 & 762 & 7754 & 10.58 & 107.69 & 10.18 \\ 
School of Rural and Surveying Engineering & 22 & 207 & 1817 & 9.41 & 82.59 & 8.78 \\ 
School of Civil Engineering & 51 & 576 & 3869 & 11.29 & 75.86 & 6.72 \\ 
School of Spatial Planning and Development & 19 & 167 & 1378 & 8.79 & 72.53 & 8.25 \\ 
School of Forestry and Natural Environment & 32 & 283 & 2119 & 8.84 & 66.22 & 7.49 \\ 
School of Dentistry & 41 & 294 & 2611 & 7.17 & 63.68 & 8.88 \\ 
School of Physical Education and Sports Science (Serres) & 24 & 190 & 1474 & 7.92 & 61.42 & 7.76 \\ 
School of Political Sciences & 19 & 78 & 900 & 4.11 & 47.37 & 11.54 \\ 
School of Economics & 28 & 160 & 1125 & 5.71 & 40.18 & 7.03 \\ 
School of Physical Education and Sports Science & 47 & 286 & 1807 & 6.09 & 38.45 & 6.32 \\ 
School of Journalism and Mass Media Studies & 22 & 126 & 723 & 5.73 & 32.86 & 5.74 \\ 
School of Mathematics & 21 & 141 & 453 & 6.71 & 21.57 & 3.21 \\ 
School of Film & 20 & 52 & 389 & 2.60 & 19.45 & 7.48 \\ 
School of English Language and Literature & 21 & 71 & 330 & 3.38 & 15.71 & 4.65 \\ 
School of Philosophy and Education & 23 & 59 & 344 & 2.57 & 14.96 & 5.83 \\ 
School of History and Archaeology & 34 & 72 & 404 & 2.12 & 11.88 & 5.61 \\ 
School of Early Childhood Education & 23 & 78 & 189 & 3.39 & 8.22 & 2.42 \\ 
School of Architecture & 25 & 28 & 153 & 1.12 & 6.12 & 5.46 \\ 
School of Primary Education & 27 & 40 & 133 & 1.48 & 4.93 & 3.33 \\ 
School of Italian Language and Literature & 14 & 26 & 60 & 1.86 & 4.29 & 2.31 \\ 
School of Music Studies & 18 & 26 & 52 & 1.44 & 2.89 & 2 \\ 
School of German Language and Literature & 17 & 13 & 41 & 0.76 & 2.41 & 3.15 \\ 
School of Philology & 55 & 64 & 61 & 1.16 & 1.11 & 0.95 \\ 
School of French Language and Literature & 17 & 10 & 7 & 0.59 & 0.41 & 0.70 \\ 
School of Law & 75 & 23 & 24 & 0.31 & 0.32 & 1.04 \\ 
School of Theology & 31 & 15 & 9 & 0.48 & 0.29 & 0.60 \\ 
School of Social Theology and Christian Culture & 31 & 15 & 9 & 0.48 & 0.29 & 0.60 \\ 
School of Drama & 17 & 5 & 2 & 0.29 & 0.12 & 0.40 \\ 
School of Visual and Applied Arts & 22 & 1 & 0 & 0.05 & 0 & 0 \\ 
\midrule
  \multicolumn{7}{l}{NKUA} \\
\midrule
Department of Physics & 57 & 2213 & 54931 & 38.82 & 963.70 & 24.82 \\ 
School of Medicine & 524 & 13565 & 265497 & 25.89 & 506.67 & 19.57 \\ 
Department of Informatics and Telecommunications & 33 & 818 & 10090 & 24.79 & 305.76 & 12.33 \\ 
Department of Pharmacy & 42 & 846 & 11992 & 20.14 & 285.52 & 14.17 \\ 
Department of Chemistry & 42 & 864 & 11683 & 20.57 & 278.17 & 13.52 \\ 
Department of Biology & 39 & 710 & 10675 & 18.21 & 273.72 & 15.04 \\ 
Department of Agricultural Development, Agrofood and Management of Natural Resources & 13 & 205 & 3522 & 15.77 & 270.92 & 17.18 \\ 
School of Nursing & 29 & 442 & 4702 & 15.24 & 162.14 & 10.64 \\ 
Department of Geology and Geoenvironment & 42 & 592 & 4862 & 14.10 & 115.76 & 8.21 \\ 
Department of Psychology & 19 & 218 & 2114 & 11.47 & 111.26 & 9.70 \\ 
Department of Port Management and Shipping & 10 & 106 & 999 & 10.60 & 99.90 & 9.42 \\ 
Department of Aerospace Science and Technology & 8 & 129 & 685 & 16.12 & 85.62 & 5.31 \\ 
School of Dentistry & 67 & 469 & 4783 & 7 & 71.39 & 10.20 \\ 
School of Physical Education and Sport Science & 49 & 351 & 2564 & 7.16 & 52.33 & 7.30 \\ 
Department of Economics & 35 & 241 & 1699 & 6.89 & 48.54 & 7.05 \\ 
Department of Primary Education & 25 & 125 & 990 & 5 & 39.60 & 7.92 \\ 
Department of Business Administration & 9 & 46 & 178 & 5.11 & 19.78 & 3.87 \\ 
Department of Mathematics & 42 & 241 & 674 & 5.74 & 16.05 & 2.80 \\ 
Department of Digital Arts and Cinema & 9 & 29 & 134 & 3.22 & 14.89 & 4.62 \\ 
Department of Sociology & 3 & 10 & 42 & 3.33 & 14 & 4.20 \\ 
Department of Educational Studies & 16 & 48 & 219 & 3 & 13.69 & 4.56 \\ 
Department of Early Childhood Education & 18 & 55 & 232 & 3.06 & 12.89 & 4.22 \\ 
Department of History and Philosophy of Science & 24 & 114 & 275 & 4.75 & 11.46 & 2.41 \\ 
Department of English Language and Literature & 23 & 96 & 259 & 4.17 & 11.26 & 2.70 \\ 
Department of Communication and Media Studies & 21 & 66 & 208 & 3.14 & 9.90 & 3.15 \\ 
Department of Political Science and Public Administration & 33 & 89 & 231 & 2.70 & 7 & 2.60 \\ 
Department of Philology & 50 & 81 & 160 & 1.62 & 3.20 & 1.98 \\ 
Department of Music Studies & 24 & 37 & 63 & 1.54 & 2.62 & 1.70 \\ 
Department of German Language and Literature & 20 & 21 & 39 & 1.05 & 1.95 & 1.86 \\ 
Department of History and Archaeology & 36 & 48 & 36 & 1.33 & 1 & 0.75 \\ 
Department of French Language and Literature & 22 & 5 & 18 & 0.23 & 0.82 & 3.60 \\ 
School of Law & 91 & 24 & 52 & 0.26 & 0.57 & 2.17 \\ 
Department of Philosophy & 13 & 11 & 7 & 0.85 & 0.54 & 0.64 \\ 
Department of Theology & 23 & 13 & 10 & 0.57 & 0.43 & 0.77 \\ 
Department of Russian Language and Literature and Slavic Studies & 7 & 17 & 1 & 2.43 & 0.14 & 0.06 \\ 
Department of Social Theology and the Study of Religion & 19 & 2 & 2 & 0.11 & 0.11 & 1 \\ 
Department of Turkish Studies and Modern Asian Studies & 10 & 3 & 0 & 0.30 & 0 & 0 \\ 
Department of Theatre Studies & 17 & 7 & 0 & 0.41 & 0 & 0 \\ 
Department of Spanish Language and Literature & 9 & 1 & 0 & 0.11 & 0 & 0 \\ 
Department of Italian Language and Literature & 11 & 2 & 0 & 0.18 & 0 & 0 \\ 
\midrule
 \multicolumn{7}{l}{UThessaly} \\
\midrule
Department of Food Science and Nutrition & 10 & 206 & 13716 & 20.60 & 1371.60 & 66.58 \\ 
Department of Physics & 13 & 533 & 17568 & 41 & 1351.38 & 32.96 \\ 
Department of Medicine & 109 & 2332 & 51473 & 21.39 & 472.23 & 22.07 \\ 
Department of Agriculture, Crop Production and Rural Environment & 24 & 645 & 8360 & 26.88 & 348.33 & 12.96 \\ 
Department of Dietetics and Nutrition & 9 & 217 & 2323 & 24.11 & 258.11 & 10.71 \\ 
Department of Informatics and Telecommunications & 16 & 415 & 3830 & 25.94 & 239.38 & 9.23 \\ 
Department of Mechanical Engineering & 22 & 351 & 4998 & 15.95 & 227.18 & 14.24 \\ 
Department of Biochemistry and Biotechnology & 24 & 317 & 5036 & 13.21 & 209.83 & 15.89 \\ 
Department of Computer Science and Biomedical Informatics & 20 & 448 & 3940 & 22.40 & 197 & 8.79 \\ 
Department of Nursing & 9 & 203 & 1728 & 22.56 & 192 & 8.51 \\ 
Department of Electrical and Computer Engineering & 24 & 630 & 4256 & 26.25 & 177.33 & 6.76 \\ 
Department of Physical Education and Sport Science & 26 & 354 & 3744 & 13.62 & 144 & 10.58 \\ 
Department of Public and One Health & 5 & 110 & 710 & 22 & 142 & 6.45 \\ 
Department of Agriculture, Ichthyology and Aquatic Environment & 10 & 131 & 1295 & 13.10 & 129.50 & 9.89 \\ 
Department of Energy Systems & 10 & 122 & 1288 & 12.20 & 128.80 & 10.56 \\ 
Department of Economics & 22 & 284 & 2829 & 12.91 & 128.59 & 9.96 \\ 
Department of Civil Engineering & 20 & 257 & 2095 & 12.85 & 104.75 & 8.15 \\ 
Department of Forestry, Wood Science and Design & 17 & 127 & 1313 & 7.47 & 77.24 & 10.34 \\ 
Department of Animal Science & 9 & 98 & 625 & 10.89 & 69.44 & 6.38 \\ 
Department of Agrotechnology & 17 & 119 & 1165 & 7 & 68.53 & 9.79 \\ 
Department of Accounting and Finance & 11 & 86 & 733 & 7.82 & 66.64 & 8.52 \\ 
Department of Mathematics & 5 & 59 & 250 & 11.80 & 50 & 4.24 \\ 
Department of Business Administration & 15 & 150 & 747 & 10 & 49.80 & 4.98 \\ 
Department of Veterinary Medicine & 36 & 329 & 1672 & 9.14 & 46.44 & 5.08 \\ 
Department of Environment & 8 & 36 & 327 & 4.50 & 40.88 & 9.08 \\ 
Department of Planning and Regional Development & 19 & 128 & 726 & 6.74 & 38.21 & 5.67 \\ 
Department of Digital Systems & 14 & 107 & 475 & 7.64 & 33.93 & 4.44 \\ 
Department of Physiotherapy & 13 & 54 & 430 & 4.15 & 33.08 & 7.96 \\ 
Department of Special Education & 20 & 79 & 456 & 3.95 & 22.80 & 5.77 \\ 
Department of Primary Education & 17 & 49 & 371 & 2.88 & 21.82 & 7.57 \\ 
Department of Architecture & 26 & 52 & 435 & 2 & 16.73 & 8.37 \\ 
General Department of Larissa & 16 & 43 & 236 & 2.69 & 14.75 & 5.49 \\ 
Department of Early Childhood Education & 21 & 59 & 266 & 2.81 & 12.67 & 4.51 \\ 
Department of History, Archaeology and Social Anthropology & 22 & 28 & 104 & 1.27 & 4.73 & 3.71 \\ 
Department of Language and Intercultural Studies & 3 & 2 & 3 & 0.67 & 1 & 1.50 \\ 
Department of Culture, Creative Media and Industries & 5 & 3 & 1 & 0.60 & 0.20 & 0.33 \\ 
\midrule
\multicolumn{7}{l}{UPatras} \\
\midrule
Department of Chemical Engineering & 25 & 657 & 11058 & 26.28 & 442.32 & 16.83 \\ 
Department of Food Science and Technology & 4 & 111 & 1360 & 27.75 & 340 & 12.25 \\ 
Faculty of Medicine & 63 & 1323 & 20064 & 21 & 318.48 & 15.17 \\ 
Department of Environmental Engineering & 10 & 197 & 3093 & 19.70 & 309.30 & 15.70 \\ 
Department of Pharmacy & 21 & 436 & 5272 & 20.76 & 251.05 & 12.09 \\ 
Department of Materials Science & 19 & 404 & 3937 & 21.26 & 207.21 & 9.75 \\ 
Department of Mechanical Engineering and Aeronautics & 29 & 612 & 5675 & 21.10 & 195.69 & 9.27 \\ 
Department of Geology & 21 & 330 & 3915 & 15.71 & 186.43 & 11.86 \\ 
Department of Chemistry & 30 & 335 & 5488 & 11.17 & 182.93 & 16.38 \\ 
Department of Physics & 27 & 446 & 4901 & 16.52 & 181.52 & 10.99 \\ 
Department of Biology & 23 & 266 & 3845 & 11.57 & 167.17 & 14.45 \\ 
Department of Civil Engineering & 21 & 268 & 3264 & 12.76 & 155.43 & 12.18 \\ 
Department of Computer Engineering and Informatics & 31 & 728 & 4195 & 23.48 & 135.32 & 5.76 \\ 
Department of Electrical and Computer Engineering & 39 & 821 & 5071 & 21.05 & 130.03 & 6.18 \\ 
Department of Economics & 17 & 86 & 1577 & 5.06 & 92.76 & 18.34 \\ 
Department of Business Administration of Food and Agricultural Enterprises & 11 & 81 & 922 & 7.36 & 83.82 & 11.38 \\ 
Department of Mathematics & 24 & 239 & 1513 & 9.96 & 63.04 & 6.33 \\ 
Department of Speech and Language Therapy & 9 & 90 & 455 & 10 & 50.56 & 5.06 \\ 
Department of Agriculture & 8 & 54 & 387 & 6.75 & 48.38 & 7.17 \\ 
Department of Museology & 2 & 16 & 96 & 8 & 48 & 6 \\ 
Department of Biosystems and Agricultural Engineering & 2 & 17 & 88 & 8.50 & 44 & 5.18 \\ 
Department of Physiotherapy & 9 & 65 & 392 & 7.22 & 43.56 & 6.03 \\ 
Department of Animal Production, Fisheries and Aquaculture & 16 & 82 & 684 & 5.12 & 42.75 & 8.34 \\ 
Department of Business Administration & 18 & 111 & 712 & 6.17 & 39.56 & 6.41 \\ 
Department of Crop Science & 3 & 22 & 116 & 7.33 & 38.67 & 5.27 \\ 
Department of Nursing & 10 & 62 & 259 & 6.20 & 25.90 & 4.18 \\ 
Department of Educational Sciences and Early Childhood Education & 24 & 121 & 517 & 5.04 & 21.54 & 4.27 \\ 
Department of History and Archaeology & 12 & 46 & 203 & 3.83 & 16.92 & 4.41 \\ 
Department of Management Science and Technology & 18 & 51 & 241 & 2.83 & 13.39 & 4.73 \\ 
Department of Tourism Management & 13 & 23 & 174 & 1.77 & 13.38 & 7.57 \\ 
Department of Education and Social Work & 21 & 88 & 203 & 4.19 & 9.67 & 2.31 \\ 
Department of Philosophy & 15 & 25 & 33 & 1.67 & 2.20 & 1.32 \\ 
Department of Philology & 22 & 40 & 47 & 1.82 & 2.14 & 1.18 \\ 
Department of Architecture & 22 & 24 & 42 & 1.09 & 1.91 & 1.75 \\ 
Department of Theater Studies & 15 & 14 & 7 & 0.93 & 0.47 & 0.50 \\ 
\midrule
\multicolumn{7}{l}{UNIWA} \\
\midrule
Department of Public Health Policies & 26 & 217 & 3605 & 8.35 & 138.65 & 16.61 \\ 
Department of Informatics and Computer Engineering & 25 & 342 & 3155 & 13.68 & 126.20 & 9.23 \\ 
Department of Mechanical Engineering & 20 & 186 & 1838 & 9.30 & 91.90 & 9.88 \\ 
Department of Public and Community Health & 18 & 138 & 1474 & 7.67 & 81.89 & 10.68 \\ 
Department of Food Science and Technology & 13 & 107 & 836 & 8.23 & 64.31 & 7.81 \\ 
Department of Biomedical Engineering & 19 & 160 & 1185 & 8.42 & 62.37 & 7.41 \\ 
Department of Biomedical Sciences & 47 & 268 & 2392 & 5.70 & 50.89 & 8.93 \\ 
Department of Electrical and Electronic Engineering & 60 & 430 & 3052 & 7.17 & 50.87 & 7.10 \\ 
Department of Industrial Design and Production Engineering & 27 & 211 & 1275 & 7.81 & 47.22 & 6.04 \\ 
Department of Physiotherapy & 15 & 109 & 467 & 7.27 & 31.13 & 4.28 \\ 
Department of Accounting and Finance & 19 & 64 & 578 & 3.37 & 30.42 & 9.03 \\ 
Department of Naval Architecture & 15 & 45 & 413 & 3 & 27.53 & 9.18 \\ 
Department of Wine, Vine and Beverage Sciences & 10 & 29 & 269 & 2.90 & 26.90 & 9.28 \\ 
Department of Nursing & 39 & 202 & 971 & 5.18 & 24.90 & 4.81 \\ 
Department of Surveying and Geoinformatics Engineering & 16 & 83 & 392 & 5.19 & 24.50 & 4.72 \\ 
Department of Business Administration & 40 & 224 & 914 & 5.60 & 22.85 & 4.08 \\ 
Department of Civil Engineering & 23 & 80 & 512 & 3.48 & 22.26 & 6.40 \\ 
Department of Archival, Library and Information Studies & 13 & 80 & 257 & 6.15 & 19.77 & 3.21 \\ 
Department of Tourism Management & 17 & 48 & 330 & 2.82 & 19.41 & 6.88 \\ 
Department of Social Work & 10 & 47 & 153 & 4.70 & 15.30 & 3.26 \\ 
Department of Midwifery & 16 & 130 & 231 & 8.12 & 14.44 & 1.78 \\ 
Department of Early Childhood Care and Education & 15 & 40 & 206 & 2.67 & 13.73 & 5.15 \\ 
Department of Occupational Therapy & 5 & 8 & 42 & 1.60 & 8.40 & 5.25 \\ 
Department of Conservation of Antiquities and Works of Art & 23 & 42 & 161 & 1.83 & 7 & 3.83 \\ 
Department of Interior Architecture & 17 & 17 & 93 & 1 & 5.47 & 5.47 \\ 
Department of Graphic Design and Visual Communication & 31 & 12 & 49 & 0.39 & 1.58 & 4.08 \\ 
Department of Photography and Audiovisual Arts & 20 & 1 & 0 & 0.05 & 0 & 0 \\ 
\midrule
 \multicolumn{7}{l}{UoI} \\
\midrule
Department of Physics & 39 & 1066 & 24134 & 27.33 & 618.82 & 22.64 \\ 
Department of Chemistry & 35 & 643 & 11338 & 18.37 & 323.94 & 17.63 \\ 
Department of Materials Science and Engineering & 22 & 665 & 6789 & 30.23 & 308.59 & 10.21 \\ 
Faculty of Medicine & 129 & 2363 & 38724 & 18.32 & 300.19 & 16.39 \\ 
Department of Agricultural Technology & 17 & 136 & 3562 & 8 & 209.53 & 26.19 \\ 
Department of Biological Applications and Technology & 20 & 205 & 2644 & 10.25 & 132.20 & 12.90 \\ 
Department of Computer Science and Engineering & 26 & 383 & 2944 & 14.73 & 113.23 & 7.69 \\ 
Department of Psychology & 8 & 45 & 876 & 5.62 & 109.50 & 19.47 \\ 
Department of Primary Education & 19 & 153 & 1717 & 8.05 & 90.37 & 11.22 \\ 
Department of Informatics and Telecommunications & 18 & 203 & 1147 & 11.28 & 63.72 & 5.65 \\ 
Department of Speech and Language Therapy & 10 & 94 & 483 & 9.40 & 48.30 & 5.14 \\ 
Department of Economics & 19 & 107 & 740 & 5.63 & 38.95 & 6.92 \\ 
Department of Mathematics & 19 & 165 & 667 & 8.68 & 35.11 & 4.04 \\ 
Department of Early Years Learning and Care & 7 & 37 & 215 & 5.29 & 30.71 & 5.81 \\ 
Department of Nursing & 7 & 32 & 155 & 4.57 & 22.14 & 4.84 \\ 
Department of Philosophy & 13 & 20 & 122 & 1.54 & 9.38 & 6.10 \\ 
Department of Early Childhood Education & 18 & 56 & 131 & 3.11 & 7.28 & 2.34 \\ 
Department of Architecture & 8 & 23 & 39 & 2.88 & 4.88 & 1.70 \\ 
Department of Accounting and Finance & 17 & 28 & 48 & 1.65 & 2.82 & 1.71 \\ 
Department of History and Archaeology & 25 & 29 & 43 & 1.16 & 1.72 & 1.48 \\ 
Department of Philology & 18 & 19 & 16 & 1.06 & 0.89 & 0.84 \\ 
Department of Music Studies & 12 & 2 & 7 & 0.17 & 0.58 & 3.50 \\ 
Department of Fine Arts and Art Sciences & 20 & 8 & 1 & 0.40 & 0.05 & 0.12 \\ 
\midrule
\multicolumn{7}{l}{DUTH} \\
\midrule
Department of Medicine & 92 & 1395 & 23297 & 15.16 & 253.23 & 16.70 \\ 
Department of Environmental Engineering & 20 & 323 & 4925 & 16.15 & 246.25 & 15.25 \\ 
Department of Agricultural Development & 17 & 249 & 3723 & 14.65 & 219 & 14.95 \\ 
Department of Forestry and Management of the Environment and of Natural Resources & 19 & 251 & 2652 & 13.21 & 139.58 & 10.57 \\ 
Department of Molecular Biology and Genetics & 23 & 223 & 3081 & 9.70 & 133.96 & 13.82 \\ 
Department of Production and Management Engineering & 15 & 206 & 1917 & 13.73 & 127.80 & 9.31 \\ 
Department of Electrical and Computer Engineering & 35 & 652 & 4466 & 18.63 & 127.60 & 6.85 \\ 
Department of Civil Engineering & 39 & 585 & 3357 & 15 & 86.08 & 5.74 \\ 
Department of Primary Education & 20 & 97 & 909 & 4.85 & 45.45 & 9.37 \\ 
Department of Economics Sciences & 19 & 95 & 586 & 5 & 30.84 & 6.17 \\ 
Department of Physical Education and Sport Science & 41 & 185 & 973 & 4.51 & 23.73 & 5.26 \\ 
Department of Social Work & 14 & 31 & 143 & 2.21 & 10.21 & 4.61 \\ 
Department of History and Ethnology & 21 & 51 & 214 & 2.43 & 10.19 & 4.20 \\ 
Department of Education Sciences in Early Childhood & 15 & 26 & 66 & 1.73 & 4.40 & 2.54 \\ 
Department of Social Policy & 7 & 12 & 28 & 1.71 & 4 & 2.33 \\ 
Department of Architectural Engineering & 18 & 7 & 69 & 0.39 & 3.83 & 9.86 \\ 
Department of Languages, Literature and Culture of Black Sea Countries & 13 & 5 & 17 & 0.38 & 1.31 & 3.40 \\ 
Department of Greek & 22 & 32 & 17 & 1.45 & 0.77 & 0.53 \\ 
Department of Political Science & 4 & 2 & 1 & 0.50 & 0.25 & 0.50 \\ 
Department of Law & 56 & 2 & 0 & 0.04 & 0 & 0 \\ 
\midrule
\multicolumn{7}{l}{UoC} \\
\midrule
Department of Materials Science and Technology & 20 & 563 & 18600 & 28.15 & 930 & 33.04 \\ 
School of Medicine & 114 & 2682 & 64380 & 23.53 & 564.74 & 24 \\ 
Department of Chemistry & 23 & 498 & 7651 & 21.65 & 332.65 & 15.36 \\ 
Department of Computer Science & 23 & 767 & 7483 & 33.35 & 325.35 & 9.76 \\ 
Department of Physics & 17 & 355 & 5030 & 20.88 & 295.88 & 14.17 \\ 
Department of Biology & 23 & 291 & 3333 & 12.65 & 144.91 & 11.45 \\ 
Department of Psychology & 18 & 155 & 2155 & 8.61 & 119.72 & 13.90 \\ 
Department of Preschool Education & 26 & 103 & 1802 & 3.96 & 69.31 & 17.50 \\ 
Department of Economics & 25 & 153 & 1541 & 6.12 & 61.64 & 10.07 \\ 
Department of Mathematics and Applied Mathematics & 33 & 182 & 1073 & 5.52 & 32.52 & 5.90 \\ 
Department of Sociology & 22 & 59 & 394 & 2.68 & 17.91 & 6.68 \\ 
Department of Primary Education & 20 & 48 & 180 & 2.40 & 9 & 3.75 \\ 
Department of Philology & 24 & 60 & 135 & 2.50 & 5.62 & 2.25 \\ 
Department of Political Science & 16 & 17 & 88 & 1.06 & 5.50 & 5.18 \\ 
Department of History and Archaeology & 23 & 40 & 77 & 1.74 & 3.35 & 1.93 \\ 
Department of Philosophy and Social Studies & 21 & 22 & 40 & 1.05 & 1.90 & 1.82 \\ 
\midrule
\multicolumn{7}{l}{NTUA} \\
\midrule
School of Applied Mathematical and Physical Sciences & 67 & 2339 & 46887 & 34.91 & 699.81 & 20.05 \\ 
School of Mechanical Engineering & 43 & 1068 & 12165 & 24.84 & 282.91 & 11.39 \\ 
School of Chemical Engineering & 47 & 867 & 11866 & 18.45 & 252.47 & 13.69 \\ 
School of Electrical and Computer Engineering & 56 & 1917 & 13812 & 34.23 & 246.64 & 7.21 \\ 
School of Civil Engineering & 47 & 885 & 8362 & 18.83 & 177.91 & 9.45 \\ 
School of Rural, Surveying, and Geoinformatics Engineering & 31 & 504 & 5076 & 16.26 & 163.74 & 10.07 \\ 
School of Mining and Metallurgical Engineering & 33 & 351 & 2764 & 10.64 & 83.76 & 7.87 \\ 
School of Naval Architecture and Marine Engineering & 22 & 280 & 1338 & 12.73 & 60.82 & 4.78 \\ 
School of Architecture & 58 & 60 & 277 & 1.03 & 4.78 & 4.62 \\ 
\midrule
\multicolumn{7}{l}{IHU} \\
\midrule
Department of Nutritional Sciences and Dietetics & 17 & 228 & 5896 & 13.41 & 346.82 & 25.86 \\ 
Department of Computer Science & 10 & 200 & 1793 & 20 & 179.30 & 8.96 \\ 
Department of Chemistry & 22 & 237 & 3460 & 10.77 & 157.27 & 14.60 \\ 
Department of Agriculture & 22 & 546 & 2692 & 24.82 & 122.36 & 4.93 \\ 
Department of Food Science and Technology & 17 & 184 & 1776 & 10.82 & 104.47 & 9.65 \\ 
Department of Forestry \& Natural Environment & 14 & 155 & 1189 & 11.07 & 84.93 & 7.67 \\ 
Department of Information and Electronic Engineering & 28 & 179 & 1630 & 6.39 & 58.21 & 9.11 \\ 
Department of Supply Chain Management & 9 & 69 & 523 & 7.67 & 58.11 & 7.58 \\ 
Department of Biomedical Sciences & 13 & 109 & 668 & 8.38 & 51.38 & 6.13 \\ 
Department of Industrial Engineering and Management & 24 & 154 & 1109 & 6.42 & 46.21 & 7.20 \\ 
Department of Nursing & 12 & 71 & 361 & 5.92 & 30.08 & 5.08 \\ 
Department of Management Science and Technology & 17 & 111 & 495 & 6.53 & 29.12 & 4.46 \\ 
Department of Civil Engineering & 11 & 58 & 285 & 5.27 & 25.91 & 4.91 \\ 
Department of Mechanical Engineering & 13 & 58 & 292 & 4.46 & 22.46 & 5.03 \\ 
Department of Physics & 14 & 90 & 292 & 6.43 & 20.86 & 3.24 \\ 
Department of Midwifery Science & 10 & 40 & 184 & 4 & 18.40 & 4.60 \\ 
Department of Environmental Engineering & 9 & 27 & 159 & 3 & 17.67 & 5.89 \\ 
Department of Organisation Management, Marketing and Tourism & 20 & 35 & 322 & 1.75 & 16.10 & 9.20 \\ 
Department of Computer, Informatics and Telecommunications Engineering & 16 & 46 & 228 & 2.88 & 14.25 & 4.96 \\ 
Department of Economic Sciences & 10 & 54 & 135 & 5.40 & 13.50 & 2.50 \\ 
Department of Surveying and Geoinformatics Engineering & 10 & 29 & 127 & 2.90 & 12.70 & 4.38 \\ 
Department of Accounting and Finance & 13 & 71 & 134 & 5.46 & 10.31 & 1.89 \\ 
Department of Creative Design and Clothing & 5 & 15 & 48 & 3 & 9.60 & 3.20 \\ 
Department of Early Childhood Education and Care & 7 & 18 & 65 & 2.57 & 9.29 & 3.61 \\ 
Department of Library, Archive and Information Science & 9 & 19 & 68 & 2.11 & 7.56 & 3.58 \\ 
Department of Business Administration & 15 & 36 & 110 & 2.40 & 7.33 & 3.06 \\ 
Department of Accounting and Information Systems & 12 & 12 & 37 & 1 & 3.08 & 3.08 \\ 
Department of Physiotherapy & 12 & 15 & 36 & 1.25 & 3 & 2.40 \\ 
Department of Interior Architecture & 3 & 4 & 4 & 1.33 & 1.33 & 1 \\ 
Department of Agricultural Biotechnology and Oenology & 0 & 0 & 0 & 0 & 0 & 0 \\ 
\midrule
 \multicolumn{7}{l}{UAegean} \\
\midrule
Department of Financial and Management Engineering & 18 & 581 & 17606 & 32.28 & 978.11 & 30.30 \\ 
Department of Marine Sciences & 17 & 279 & 4458 & 16.41 & 262.24 & 15.98 \\ 
Department of Information and Communication Systems Engineering & 27 & 506 & 5343 & 18.74 & 197.89 & 10.56 \\ 
Department of Food Science and Nutrition & 13 & 188 & 2464 & 14.46 & 189.54 & 13.11 \\ 
Department of Environment & 20 & 246 & 3643 & 12.30 & 182.15 & 14.81 \\ 
Department of Geography & 21 & 200 & 1909 & 9.52 & 90.90 & 9.54 \\ 
Department of Cultural Technology and Communication & 17 & 212 & 1139 & 12.47 & 67 & 5.37 \\ 
Department of Product and Systems Design Engineering & 19 & 208 & 1164 & 10.95 & 61.26 & 5.60 \\ 
Department of Business Administration & 23 & 117 & 862 & 5.09 & 37.48 & 7.37 \\ 
Department of Shipping, Trade and Transport & 19 & 81 & 675 & 4.26 & 35.53 & 8.33 \\ 
Department of Sociology & 14 & 56 & 342 & 4 & 24.43 & 6.11 \\ 
Department of Statistics and Actuarial-Financial Mathematics & 12 & 89 & 226 & 7.42 & 18.83 & 2.54 \\ 
Department of Tourism Economics and Management & 5 & 31 & 90 & 6.20 & 18 & 2.90 \\ 
Department of Mathematics & 20 & 79 & 281 & 3.95 & 14.05 & 3.56 \\ 
Department of Primary Education & 16 & 60 & 187 & 3.75 & 11.69 & 3.12 \\ 
Department of Pre-school Education and Educational Design & 18 & 28 & 78 & 1.56 & 4.33 & 2.79 \\ 
Department of Mediterranean Studies & 20 & 29 & 74 & 1.45 & 3.70 & 2.55 \\ 
Department of Social Anthropology and History & 19 & 15 & 15 & 0.79 & 0.79 & 1 \\ 
\midrule
 \multicolumn{7}{l}{UoP} \\
\midrule
Department of Physiotherapy & 6 & 109 & 885 & 18.17 & 147.50 & 8.12 \\ 
Department of Nutrition and Dietetics & 2 & 35 & 270 & 17.50 & 135 & 7.71 \\ 
Department of Informatics and Telecommunications & 22 & 388 & 2814 & 17.64 & 127.91 & 7.25 \\ 
Department of Nursing & 15 & 224 & 1575 & 14.93 & 105 & 7.03 \\ 
Department of Social and Education Policy & 20 & 141 & 2018 & 7.05 & 100.90 & 14.31 \\ 
Department Of Digital Systems & 7 & 112 & 604 & 16 & 86.29 & 5.39 \\ 
Department of Food Science and Technology & 13 & 99 & 1032 & 7.62 & 79.38 & 10.42 \\ 
Department Of Management Science And Technology & 8 & 104 & 553 & 13 & 69.12 & 5.32 \\ 
Department of Electrical and Computer Engineering & 32 & 305 & 1607 & 9.53 & 50.22 & 5.27 \\ 
Department of Economics & 11 & 72 & 490 & 6.55 & 44.55 & 6.81 \\ 
Department of Speech and Language Therapy & 4 & 29 & 175 & 7.25 & 43.75 & 6.03 \\ 
Department of Mechanical Engineering & 15 & 73 & 509 & 4.87 & 33.93 & 6.97 \\ 
Department of Accounting and Finance & 11 & 60 & 372 & 5.45 & 33.82 & 6.20 \\ 
Department of Performing and Digital Arts & 6 & 21 & 125 & 3.50 & 20.83 & 5.95 \\ 
Department of Agriculture & 22 & 71 & 439 & 3.23 & 19.95 & 6.18 \\ 
Department of Civil Engineering & 10 & 43 & 161 & 4.30 & 16.10 & 3.74 \\ 
Department of Sport Management & 14 & 36 & 202 & 2.57 & 14.43 & 5.61 \\ 
Department of History, Archaeology and Cultural Resources Management & 13 & 37 & 149 & 2.85 & 11.46 & 4.03 \\ 
Department of Business and Organization Administration & 18 & 33 & 140 & 1.83 & 7.78 & 4.24 \\ 
Department of Political Science and International Relations & 17 & 43 & 83 & 2.53 & 4.88 & 1.93 \\ 
Department of Philology & 14 & 11 & 2 & 0.79 & 0.14 & 0.18 \\ 
Department of Theatre Studies & 16 & 1 & 0 & 0.06 & 0 & 0 \\ 
\midrule
\multicolumn{7}{l}{UoWM} \\
\midrule
Department of Chemical Engineering & 11 & 201 & 4089 & 18.27 & 371.73 & 20.34 \\ 
Department of Electrical and Computer Engineering & 27 & 517 & 3810 & 19.15 & 141.11 & 7.37 \\ 
Department of Economics & 9 & 116 & 673 & 12.89 & 74.78 & 5.80 \\ 
Department of Occupational Therapy & 3 & 12 & 205 & 4 & 68.33 & 17.08 \\ 
Department of Statistics and Insurance Science & 2 & 14 & 127 & 7 & 63.50 & 9.07 \\ 
Department of Mathematics & 5 & 56 & 308 & 11.20 & 61.60 & 5.50 \\ 
Department of Mechanical Engineering & 25 & 172 & 1495 & 6.88 & 59.80 & 8.69 \\ 
Department of Business Administration & 7 & 49 & 331 & 7 & 47.29 & 6.76 \\ 
Department of Regional and Cross-Border Development Studies & 7 & 58 & 330 & 8.29 & 47.14 & 5.69 \\ 
Department of Communication and Digital Media & 8 & 41 & 325 & 5.12 & 40.62 & 7.93 \\ 
Department of Informatics & 8 & 72 & 296 & 9 & 37 & 4.11 \\ 
Department of Agriculture & 11 & 71 & 404 & 6.45 & 36.73 & 5.69 \\ 
Department of Product and Systems Design Engineering & 9 & 69 & 312 & 7.67 & 34.67 & 4.52 \\ 
Department of Psychology & 2 & 15 & 51 & 7.50 & 25.50 & 3.40 \\ 
Department of Management Science and Technology & 7 & 38 & 175 & 5.43 & 25 & 4.61 \\ 
Department of Accounting and Finance & 10 & 52 & 250 & 5.20 & 25 & 4.81 \\ 
Department of Midwifery & 5 & 47 & 114 & 9.40 & 22.80 & 2.43 \\ 
Department of Mineral Resources Engineering & 9 & 46 & 174 & 5.11 & 19.33 & 3.78 \\ 
Department of Early Childhood Education & 12 & 38 & 133 & 3.17 & 11.08 & 3.50 \\ 
Department of Primary Education & 17 & 41 & 119 & 2.41 & 7 & 2.90 \\ 
Department of International and European Economic Studies & 5 & 8 & 8 & 1.60 & 1.60 & 1 \\ 
Department of Fine and Applied Arts & 11 & 6 & 2 & 0.55 & 0.18 & 0.33 \\ 
\midrule
  \multicolumn{7}{l}{AUA} \\
\midrule
  Department of Food Science and Human Nutrition & 32 & 793 & 15600 & 24.78 & 487.50 & 19.67 \\ 
  Department of Crop Science & 51 & 943 & 10062 & 18.49 & 197.29 & 10.67 \\ 
  Department of Forestry and Natural Environment Management & 10 & 68 & 1824 & 6.80 & 182.40 & 26.82 \\ 
  Department of Natural Resources Development \& Agricultural Engineering & 25 & 364 & 4255 & 14.56 & 170.20 & 11.69 \\ 
  Department of Biotechnology & 21 & 302 & 2277 & 14.38 & 108.43 & 7.54 \\ 
  Department of Regional and Economic Development & 5 & 38 & 520 & 7.60 & 104 & 13.68 \\ 
  Department of Animal Science & 25 & 251 & 2303 & 10.04 & 92.12 & 9.18 \\ 
  Department of Agricultural Economics and Development & 19 & 171 & 1377 & 9 & 72.47 & 8.05 \\ 
  Department of Agribusiness and Supply Chain Management & 11 & 99 & 397 & 9 & 36.09 & 4.01 \\ 
\midrule
  \multicolumn{7}{l}{Panteion} \\
\midrule
  Department of Economic and Regional Development & 20 & 136 & 1129 & 6.80 & 56.45 & 8.30 \\ 
  Department of Psychology & 18 & 129 & 659 & 7.17 & 36.61 & 5.11 \\ 
  Department of Social Policy & 22 & 67 & 300 & 3.05 & 13.64 & 4.48 \\ 
  Department of Social Anthropology & 18 & 24 & 149 & 1.33 & 8.28 & 6.21 \\ 
  Department of Political Science and History & 27 & 35 & 222 & 1.30 & 8.22 & 6.34 \\ 
  Department of Public Administration & 23 & 33 & 157 & 1.43 & 6.83 & 4.76 \\ 
  Department of Communication, Media and Culture & 18 & 33 & 113 & 1.83 & 6.28 & 3.42 \\ 
  Department of Sociology & 26 & 34 & 158 & 1.31 & 6.08 & 4.65 \\ 
  Department of International, European and Area Studies & 27 & 44 & 96 & 1.63 & 3.56 & 2.18 \\ 
\midrule
  \multicolumn{7}{l}{UNIPI} \\
\midrule
  Department of Digital Systems & 22 & 622 & 4378 & 28.27 & 199 & 7.04 \\ 
  Department of Informatics & 22 & 496 & 4227 & 22.55 & 192.14 & 8.52 \\ 
  Department of Banking and Financial Management & 15 & 192 & 2468 & 12.80 & 164.53 & 12.85 \\ 
  Department of Industrial Management and Technology & 16 & 133 & 1580 & 8.31 & 98.75 & 11.88 \\ 
  Department of Business Administration & 26 & 161 & 2271 & 6.19 & 87.35 & 14.11 \\ 
  Department of Maritime Studies & 17 & 155 & 1375 & 9.12 & 80.88 & 8.87 \\ 
  Department of Statistics and Insurance Science & 18 & 188 & 1005 & 10.44 & 55.83 & 5.35 \\ 
  Department of Economics & 21 & 164 & 1081 & 7.81 & 51.48 & 6.59 \\ 
  Department of International and European Studies & 24 & 104 & 918 & 4.33 & 38.25 & 8.83 \\ 
  Department of Tourism Studies & 4 & 9 & 43 & 2.25 & 10.75 & 4.78 \\ 
\midrule
  \multicolumn{7}{l}{IONIO} \\
\midrule
  Department of Informatics & 17 & 461 & 3088 & 27.12 & 181.65 & 6.70 \\ 
  Department of Environment & 17 & 226 & 3035 & 13.29 & 178.53 & 13.43 \\ 
  Department of Food Science and Technology & 10 & 77 & 1420 & 7.70 & 142 & 18.44 \\ 
  Department of Tourism & 10 & 94 & 973 & 9.40 & 97.30 & 10.35 \\ 
  Department of Digital Media and Communication & 11 & 113 & 595 & 10.27 & 54.09 & 5.27 \\ 
  Department of Regional Development & 7 & 34 & 227 & 4.86 & 32.43 & 6.68 \\ 
  Department of Archives, Library Science and Museum Studies & 15 & 57 & 175 & 3.80 & 11.67 & 3.07 \\ 
  Department of Audio and Visual Arts & 22 & 52 & 154 & 2.36 & 7 & 2.96 \\ 
  Department of History & 21 & 30 & 106 & 1.43 & 5.05 & 3.53 \\ 
  Department of Foreign Languages, Translation and Interpreting & 24 & 23 & 47 & 0.96 & 1.96 & 2.04 \\ 
  Department of Ethnomusicology & 1 & 1 & 1 & 1 & 1 & 1 \\ 
  Department of Music Studies & 26 & 8 & 2 & 0.31 & 0.08 & 0.25 \\ 
\midrule
  \multicolumn{7}{l}{AUEB} \\
\midrule
  Department of Accounting and Finance & 24 & 213 & 3477 & 8.88 & 144.88 & 16.32 \\ 
  Department of Informatics & 24 & 426 & 2875 & 17.75 & 119.79 & 6.75 \\ 
  Department of Management Science and Technology & 22 & 187 & 2383 & 8.50 & 108.32 & 12.74 \\ 
  Department of Marketing and Communication & 17 & 103 & 1586 & 6.06 & 93.29 & 15.40 \\ 
  Department of Statistics & 23 & 166 & 1117 & 7.22 & 48.57 & 6.73 \\ 
  Department of Business Administration & 26 & 138 & 1034 & 5.31 & 39.77 & 7.49 \\ 
  Department of International and European Economic Studies & 25 & 120 & 723 & 4.80 & 28.92 & 6.03 \\ 
  Department of Economics & 18 & 86 & 348 & 4.78 & 19.33 & 4.05 \\ 
\midrule
  \multicolumn{7}{l}{UoM} \\
\midrule
  Department of Business Administration & 25 & 274 & 2182 & 10.96 & 87.28 & 7.96 \\ 
  Department of Economics & 19 & 150 & 1392 & 7.89 & 73.26 & 9.28 \\ 
  Department of Applied Informatics & 5 & 46 & 294 & 9.20 & 58.80 & 6.39 \\ 
  Department of Educational and Social Policy & 20 & 109 & 695 & 5.45 & 34.75 & 6.38 \\ 
  Department of Accounting and Finance & 17 & 51 & 492 & 3 & 28.94 & 9.65 \\ 
  Department of International and European Studies & 24 & 98 & 306 & 4.08 & 12.75 & 3.12 \\ 
  Department of Balkan, Slavic and Oriental Studies & 34 & 78 & 183 & 2.29 & 5.38 & 2.35 \\ 
  Department of Music Science and Art & 27 & 10 & 10 & 0.37 & 0.37 & 1 \\ 
\midrule
  \multicolumn{7}{l}{HMU} \\
\midrule
  Department of Nutrition and Dietetics & 7 & 61 & 4353 & 8.71 & 621.86 & 71.36 \\ 
  Department of Mechanical Engineering & 18 & 228 & 3789 & 12.67 & 210.50 & 16.62 \\ 
  Department of Management Science and Technology & 9 & 188 & 1487 & 20.89 & 165.22 & 7.91 \\ 
  Department of Electrical and Computer Engineering & 36 & 464 & 4640 & 12.89 & 128.89 & 10 \\ 
  Department of Agriculture & 21 & 173 & 2414 & 8.24 & 114.95 & 13.95 \\ 
  Department of Accounting and Finance & 11 & 87 & 906 & 7.91 & 82.36 & 10.41 \\ 
  Department of Electronic Engineering & 24 & 174 & 1872 & 7.25 & 78 & 10.76 \\ 
  Department of Social Work & 11 & 69 & 618 & 6.27 & 56.18 & 8.96 \\ 
  Department of Nursing & 11 & 77 & 592 & 7 & 53.82 & 7.69 \\ 
  Department of Music Technology and Acoustics & 10 & 69 & 402 & 6.90 & 40.20 & 5.83 \\ 
  Department of Business Administration & 10 & 46 & 216 & 4.60 & 21.60 & 4.70 \\ 
\midrule
  \multicolumn{7}{l}{TUC} \\
\midrule
  School of Chemical and Environmental Engineering & 25 & 533 & 10795 & 21.32 & 431.80 & 20.25 \\ 
  School of Production Engineering and Management & 28 & 474 & 5755 & 16.93 & 205.54 & 12.14 \\ 
  School of Electrical and Computer Engineering & 26 & 538 & 5009 & 20.69 & 192.65 & 9.31 \\ 
  School of Mineral Resources Engineering & 18 & 229 & 2154 & 12.72 & 119.67 & 9.41 \\ 
  School of Architecture & 20 & 64 & 252 & 3.20 & 12.60 & 3.94 \\ 
\midrule
  \multicolumn{7}{l}{HUA} \\
\midrule
  Department of Nutrition and Dietetics & 23 & 624 & 13133 & 27.13 & 571 & 21.05 \\ 
  Department of Geography & 14 & 230 & 2495 & 16.43 & 178.21 & 10.85 \\ 
  Department of Informatics and Telematics & 18 & 342 & 2579 & 19 & 143.28 & 7.54 \\ 
  Department of Economics and Sustainable Development & 15 & 85 & 502 & 5.67 & 33.47 & 5.91 \\ 
\midrule
   \multicolumn{7}{l}{ASPAITE} \\
\midrule
   
   Department of Civil Engineering Educators & 5 & 87 & 3004 & 17.40 & 600.80 & 34.53 \\ 
   Department of Electrical and Electronic Engineering Educators & 13 & 158 & 1207 & 12.15 & 92.85 & 7.64 \\ 
   Department of Mechanical Engineering Educators & 5 & 48 & 404 & 9.60 & 80.80 & 8.42 \\ 
   Department of Education & 24 & 57 & 301 & 2.38 & 12.54 & 5.28 \\ 
\midrule
   \multicolumn{7}{l}{HOU} \\
\midrule
   School of Science and Technology & 13 & 595 & 11759 & 45.77 & 904.54 & 19.76 \\ 
   School of Applied Arts and Sustainable Design & 4 & 83 & 760 & 20.75 & 190 & 9.16 \\ 
   School of Social Sciences & 12 & 63 & 1319 & 5.25 & 109.92 & 20.94 \\ 
   School of Humanities & 15 & 18 & 15 & 1.20 & 1 & 0.83 \\ 
\midrule
   \multicolumn{7}{l}{ASFA} \\
\midrule
   Department of History and Theory of Art & 10 & 6 & 4 & 0.60 & 0.40 & 0.67 \\ 
   Department of Fine Arts & 30 & 4 & 7 & 0.13 & 0.23 & 1.75 \\ 
\bottomrule 
\end{longtable}

\end{document}